\documentclass[twocolumn]{autart}

\usepackage[utf8]{inputenc}
\usepackage{url}
\usepackage{longtable}
\usepackage{amsmath}
\usepackage{amstext}
\usepackage{amssymb}
\usepackage[main=english,ngerman]{babel}

\usepackage{psfrag}
\usepackage{epsfig}
\usepackage{graphics}

\newcommand{\begfig}[1]{
        \epsfxsize#1
        \begin{figure}[htbp]
        \leavevmode
        \psfragscanon
}
\newcommand{\efig}[4]{
        \centering
        \epsffile{#2}
        \par
        \psfragscanoff
        \vspace{0.3cm}
        \refstepcounter{figure}
        \begin{minipage}[t]{#1}{\centering{\small
        Figure \arabic{figure}:\,\,\,#3} \par}
        \end{minipage}
        \protect\label{#4}
        \end{figure}
        \addcontentsline{lof}{figure}{\numberline {\arabic{figure}}{\ignorespaces #3}}
}

\newcommand{\begfigwdt}[1]{
        \epsfxsize#1
        \begin{figure*}[htbp]
        \leavevmode
        \psfragscanon
}
\newcommand{\efigwdt}[4]{
        \centering
        \epsffile{#2}
        \par
        \psfragscanoff
        \vspace{0.3cm}
        \refstepcounter{figure}
        \begin{minipage}[t]{#1}{\centering{\small
        Figure \arabic{figure}:\,\,\,#3} \par}
        \end{minipage}
        \protect\label{#4}
        \end{figure*}
        \addcontentsline{lof}{figure}{\numberline {\arabic{figure}}{\ignorespaces #3}}
}

\newcommand{\efigoln}[4]{
        \centering
        \epsffile{#2}
        \par
        \psfragscanoff
        \vspace{0.3cm}
        \refstepcounter{figure}
        \begin{minipage}[t]{#1}{\centering{\small
        Bild \arabic{figure}:\,\,\,#3} \par}
        \end{minipage}
        \protect\label{#4}
        \end{figure}
        \addcontentsline{lof}{figure}{\numberline {\arabic{figure}}{\ignorespaces #3}}
}

\newcommand{\efigap}[4]{
        \centering
        \epsffile{#2}
        \par
        \psfragscanoff
        \vspace{0.3cm}
        \refstepcounter{figure}
        \begin{minipage}[t]{#1}{\centering{\small
        Figure \Alph{section}.\arabic{figure}:\,\,\,#3} \par}
        \end{minipage}
        \protect\label{#4}
        \end{figure}
        \addcontentsline{lof}{figure}{\numberline {\Alph{section}.\arabic{figure}}{\ignorespaces #3}}
}

\newcommand{\etwofig}[6]{
        \leavevmode
        \begin{minipage}[htbp]{2#1}
        \begin{minipage}[htbp]{#1}
          \epsffile{#2}
        \end{minipage} \hspace{#4}
        \begin{minipage}[htbp]{#1}
          \epsffile{#3}
        \end{minipage}
        \end{minipage}
        \par
        \psfragscanoff
        \vspace{0.3cm}
        \refstepcounter{figure}
        \begin{minipage}[t]{15cm}{\centering{\small
        Figure \arabic{figure}:\,\,\,#5} \par}
        \end{minipage}
        \protect\label{#6}
        \end{figure}
        \addcontentsline{lof}{figure}{\numberline {\arabic{figure}}{\ignorespaces #5}}
}

\newcommand{\ethreefig}[7]{
        \hspace{-#5} \leavevmode
        \begin{minipage}[htbp]{3#1}
        \begin{minipage}[htbp]{#1}
          \epsffile{#2}
        \end{minipage}
        \begin{minipage}[htbp]{#1}
          \epsffile{#3}
        \end{minipage}
        \begin{minipage}[htbp]{#1}
          \epsffile{#4}
        \end{minipage}
        \end{minipage} \par
        \psfragscanoff
        \vspace{0.3cm}
        \refstepcounter{figure}
        \begin{minipage}[t]{15cm}{\centering{\small
        Figure \arabic{figure}:\,\,\,#6} \par}
        \end{minipage}
        \protect\label{#7}
        \end{figure}
        \addcontentsline{lof}{figure}{\numberline {\arabic{figure}}{\ignorespaces #6}}
}

\newcommand{\efourfig}[8]{
        \hspace{#6}\leavevmode
        \begin{minipage}[htbp]{2#1}
            \begin{minipage}[htbp]{#1}
             \epsffile{#2}
            \end{minipage}
            \begin{minipage}[htbp]{#1}
             \epsffile{#3}
            \end{minipage} \\[4ex]
            \begin{minipage}[htbp]{#1}
             \epsffile{#4}
            \end{minipage}
            \begin{minipage}[htbp]{#1}
             \epsffile{#5}
            \end{minipage}
        \end{minipage} 
        \par
        \psfragscanoff
        \vspace{0.3cm}
        \refstepcounter{figure}
        \begin{minipage}[t]{15cm}{\centering{\small
        Figure \arabic{figure}:\,\,\,#7} \par}
        \end{minipage}
        \protect\label{#8}
   \end{figure}
   \addcontentsline{lof}{figure}{\numberline {\arabic{figure}}{\ignorespaces #7}}
 }

\newcommand{\ind}[2]{ \indent{ \begin{tabular}{#1} #2 \end{tabular}} }

\newcommand{\tabcap}[6]{
\begin{table}[#1]
\centering
\begin{tabular}{#2} #3 \end{tabular} \refstepcounter{table} \label{#6}
\\[1.5ex]
\par {\bmin{#4} \centering Table \arabic{table}: #5 \emin} \par
\addcontentsline{lot}{table}{\numberline {\arabic{table}}{\ignorespaces #5}}
\end{table}
}

\newcommand{\bmin}[2]{ \begin{minipage}[htbp]{#1} #2}
\newcommand{\emin}{ \end{minipage}}

\newtheorem{TaskT}{Task}[section]

\newtheorem{ProblemT}{Problem}[section]

\newtheorem{PropertyT}{Property}[section]
\newcommand{\Property}[2]
{       \begin{PropertyT}
        #1 $\Box$
        \label{#2}
        \end{PropertyT}
}

\newtheorem{SolutionT}{Solution}[section]

\newtheorem{AssumptionT}{Assumption}[section]

\newtheorem{DefinitionT}{Definition}[section]
\newcommand{\Definition}[2]
{       \begin{DefinitionT}
        #1 $\Box$
        \label{#2}
        \end{DefinitionT}
}

\newtheorem{AlgorithmT}{Algorithm}[section]

\newtheorem{CorollaryT}{Corollary}[section]

\newtheorem{LemmaT}{Lemma}[section]

\newtheorem{ProofT}{Proof}[section]

\newcommand{\be}{\begin{equation}}
\newcommand{\ee}{\end{equation}}

\newcommand{\ba}{\begin{array}{*{20}c}}
\newcommand{\baa}[1]{\begin{array}{*{20}#1}}
\newcommand{\bega}[1]{\begin{array}{#1}}
\newcommand{\ea}{\end{array}}

\newcommand{\bab}{\left( \begin{array}{*{20}c}}
\newcommand{\baab}[1]{\left( \begin{array}{*{20}#1}}
\newcommand{\begab}[1]{\left( \begin{array}{#1}}
\newcommand{\eab}{\end{array} \right)}

\def\Mbf#1{\mbox{\boldmath$#1$}}
\newcommand{\dv}{\mathsf v}

\newcommand{\dd}{\mathsf d}
\newcommand{\dw}{\mathsf w}
\newcommand{\de}{\mathsf e}
\newcommand{\vekdd}{\Mbf {\mathsf d}}
\newcommand{\vekde}{\Mbf {\mathsf e}}
\newcommand{\vekdv}{\Mbf {\mathsf v}}
\newcommand{\vekdw}{\Mbf {\mathsf w}}
\newcommand{\vekdz}{\Mbf {\mathsf z}}
\newcommand{\matA}{\Mbf A}

\newcommand{\vekf}{\Mbf f}

\newcommand{\matI}{\Mbf I}

\newcommand{\vekx}{\Mbf x}
\newcommand{\vekxdot}{\dot{\Mbf x}}
\newcommand{\veku}{\Mbf u}
\newcommand{\vekd}{\Mbf d}

\newcommand{\vekz}{\Mbf z}

\begin{document} 

\begin{frontmatter}

\title{Flat Hybrid Automata as a Class of Reachable Systems: Introductory Theory and Examples}

\author[tk]{Tobias Kleinert}\ead{tobias.kleinert@basf.com},
\author[vh]{Veit Hagenmeyer}\ead{veit.hagenmeyer@kit.edu}

\address[tk]{Technical Site Services Automation, BASF Schwarzheide GmbH, Schwarzheide, Germany}  
\address[vh]{Institute for Automation and Applied Informatics, Karlsruhe Institute of Technology, Karlsruhe, Germany}


\begin{abstract}                          
Controlling hybrid systems is mostly very challenging due to the variety of dynamics these systems can exhibit. Inspired by the concept of differential flatness of nonlinear continuous systems and their inherent invertibility property, the present contribution is focused on explicit input trajectory calculation. To this end, a new class of hybrid systems called Flat Hybrid Automata is introduced as a realisation of deterministic, reachable and explicitly invertible hybrid automata. Relevant system properties are derived, an approach for  construction and for trajectory calculation is proposed and two demonstrative examples are presented. The results constitute a generalisation of control of invertible hybrid systems which is very useful if, e.g., fast reaction for stabilisation or transitions is relevant. 
\end{abstract}

\end{frontmatter}



\renewcommand{\baselinestretch}{1.4}



\section{Introduction}
\label{s:intro}

Discrete and continuous control is evenly relevant in practical applications. Its implementation is typically "hybrid" i.e., separated into interacting discrete and continuous parts, an approach that allows to systematically formulate and solve the control task. However, the resulting systems can exhibit considerable combinatory complexity and non--deterministic dynamical behaviour \cite{AlCoHeHo93,KoGaetal2009,LuLa2009,ScAbMaetal2015}. Inspired by the concept of differential flatness of nonlinear continuous systems and their inherent invertibility property, the present contribution is focused on hybrid automata with input and output, the discrete and continuous input trajectories of which can be determined explicitly from system inversion given the output trajectories. Thereby, handling of the typical hybrid system's complexity can be avoided. The new system class is called Flat Hybrid Automaton ($\rm FHA$), a hybrid automaton consisting of the discrete--event subsystem $\rm A$, the continuous--valued, continuous--time subsystem $\rm C$ and a set of deterministic continuous and discrete switching rules which interlink the two subsystems. It is supposed that $\rm A$ is deterministic and strongly connected (see, e.g., \cite{BrePle1979}), $\rm C$ is differentially flat (see, e.g., \cite{FlLeMaRo1992}), and the continuous switching rules are defined on the flat output of $\rm C$, only. It is shown that, if these properties are given, then all discrete states and continuous outputs are reachable and  $\rm A$ and $\rm C$ are invertible in the sense that the continuous and discrete input trajectories are explicitly determinable from sequences of switching rules for given discrete output sequences and given continuous initial and target conditions. 

The paper is organized as follows: Related literature is reviewed in Section \ref{s:literature}. System definition, description of central system properties and construction as well as trajectory planning are developed in Section \ref{s:fha} and \ref{s:constr_trajplan}, respectively.  Demonstrative examples of a tank system and an electrical network are described in Section \ref{s:fha_examples}. The paper is concluded and an outlook is given in Section \ref{s:concl_outl}.


\section{Related work}
\label{s:literature}

The work presented is based on hybrid automata \cite{AlCoHaHeHoNiOlSiYo1995,AlCoHeHo93,HedeSchuLuLa2010}, differential flatness \cite{FlLeMaRo1992,FlLeMaRo1995,SiRaAg2004}, matrix analysis and graph theory \cite{BrePle1979,Ga1959,Me2000}. In general, the work can in parts be considered as a hierarchical control system (cf., e.g., \cite{RaYo1998}). First publications on differential flatness in connection with discrete--event systems date back to about the year 2000. The publications can be grouped into application--related, with focus on simplification of control design using differential flatness, and more conceptually oriented work considering system theoretical questions: 

In application--oriented works like e.g., \cite{GeWoRuGue2006,RoSiRa:2003,SiRaSiOr:2002}, design of input--output linearisation and feed--forward trajectory calculation is addressed for switching systems of which the continuous subsystems are differentially flat. The results show that, if applicable, differential flatness can significantly contribute to simplifying control design. In \cite{DrGuMi2018,MiDa2007,MiDa2009,PaMi2013}, system inversion and flatness is explicitly addressed in the context of secure communications, with application to linear discrete--time systems that are subject to externally triggered switching. Conditions for system inversion are investigated and derived. The central intend is to reconstruct continuous input signals by applying system inversion. Parts of these results can be applied for planning transition control of linear switched discrete-time continuous systems. 

An interesting trace between flatness and hybrid systems can be found in the more system theoretical oriented work of Paulo Tabuada and co-workers. In \cite{tabuada:2004}, the notion of flatness is related to transition systems in the context of bisimulation. It is shown that finite bisimulation systems can be constructed for differentially flat nonlinear discrete--time systems. In \cite{tab_pap_lim:2004}, a class of general control systems capturing both continuous--valued and discrete--event systems as well as hybrid systems with both continuous and discrete inputs is described. Towards controlling such systems, model abstraction, bisimulation and composition of abstract control systems is developed. This consideration of flatness in hybrid systems has a relevance regarding system theoretical development of bisimulation in systems control. 

On this background, the present paper addresses in a new way the inversion of hybrid dynamical systems, in order to establish deterministic dynamical behaviour and reachability as well as to explicitly determine control input trajectories. Thereby, methods for system construction and trajectory planning are provided in view of technically relevant systems. For taming complexity, a strong emphasis is put on the aspect of designing the to-be-controlled system such that it is flat.


\section{The Flat Hybrid Automaton}
\label{s:fha}

\subsection{Hybrid systems' variables}
\label{ss:hybrd_dyn}

Representing the sub--dynamics of a hybrid system by an automaton and continuous--time state--space models yields a hybrid automaton as introduced in \cite{AlCoHaHeHoNiOlSiYo1995,AlCoHeHo93}. In general, the sub--systems can exhibit various kinds of dynamics. For the introduction of the Flat Hybrid Automaton, discrete and continuous subsystems with input, output and deterministic dynamical behaviour are considered in the sense that given the initial state and an input trajectory, the state and output trajectories exist and are unique. The continuous subsystems $\rm C$ are represented by continuous--time nonlinear state--space models. The notation used in the following is based on the one developed in \cite{Kl06b} and \cite{KlLu02}. \\[-4ex]

System variables of the discrete subsystem $\rm A$ are {\it discrete states} \\[-1.5ex]
\begin{equation}\nonumber
\dd_i \in \{0,1\} \,,\; i=1, \dots , n\dd 
\end{equation}
with $\vekdd$ the $n\dd$--dimensional vector of discrete states $\dd_i$, {\it discrete inputs}  \\[-1.5ex]
\begin{equation}\nonumber
\dv_i \in \{0,1\} \,,\; i=1, \dots , n\dv 
\end{equation}
with $\vekdv$ the $n\dv$--dimensional vector of the discrete inputs $\dv_i$, and {\it discrete outputs}  \\[-1.5ex]
\begin{equation}\nonumber
\dw_i \in \{0,1\} \,,\; i=1, \dots , n\dw 
\end{equation}
with $\vekdw$ the $n\dw$--dimensional vector of discrete outputs $\dw_i$. If a $\dd$, $\dv$ or $\dw$ equals $1$, it is considered active, else inactive. \\[-4ex]

The continuous sub--system variables are {\it vectors of continuous states} \\[-1.5ex]
\begin{equation}\nonumber
\vekx_{\dd i} \in \mathcal{X}_{\dd i}
\end{equation}
of the continuous--state space $\mathcal{X}_{\dd i}$, bi--uniquely assigned to a $\dd_i$, {\it vectors of continuous inputs} \\[-1.5ex]
\begin{equation}\nonumber
\veku_{\dd i} \in \mathcal{U}_{\dd i}
\end{equation}
of the continuous--input space $\mathcal{U}_{\dd i}$, bi--uniquely assigned to $\dd_i$ and {\it vectors of continuous outputs} \\[-1.5ex]
\begin{equation}\nonumber
\vekz_{\dd i} \in \mathcal{Z}_{\dd i}
\end{equation}
of the continuous--output space $\mathcal{Z}_{\dd i}$, also bi--uniquely assigned to $\dd_i$. \\[-4ex]

It is suitable to represent the evolution of the {\it discrete--state trajectory} \\[-1.5ex] 
\begin{equation}\nonumber
\vekdd = \vekdd(k) 
\end{equation}
using $k \in \mathbb{N}_0$ as discrete time variable counting events. A step of $k$ indicates that the discrete state has changed, which in the following is called {\it discrete--state switching} i.e., \\[-1.5ex]
\begin{equation}\nonumber
\dd_{i1} \rightarrow \dd_{i2} \,, \; i_1, i_2 \in \{ 1, \dots , n\dd \} \,, \; i_1 \neq i_2 \,,
\end{equation}
with $\dd_{i2} = \dd_{i2}(k+1) = \dd_{i1}'$, the discrete successor state of $\dd_{i1} = \dd_{i1}(k)$, the discrete predecessor state, and $\vekdd'$ the successor of $\vekdd$, respectively. For a constant $k$, only one $\dd_i$ may be active. A change of the continuous state, during a discrete--state switching, is called {\it continuous--state switching} and $\vekx'$ is the continuous successor state of $\vekx$. The combined event \\[-1.5ex]
\begin{equation}\nonumber
(\vekdd,\vekx) \rightarrow (\vekdd',\vekx') 
\end{equation}
is called {\it state switching} in the following. \\[-4ex]

In order to link discrete--state switching to continuous time $t \in {\mathbb R}_0^+$, it is useful to relate $k$ to $t$: $k = k(t)$. The instant when a discrete--state switching has taken place is commonly denoted with $t'$, in accordance to the notation of the successor state $(\vekdd',\vekx')$. Theoretically, several discrete--state switching can occur at the same time point $t$, in form of a switching sequence of duration $0$. It is, furthermore, assumed, in a first approach, that discrete as well as continuous inputs can be set at any time $t$ to any value of their input spaces. The following representation of time--dependence of the state, input and output is obtained: \\[-1.5ex]
\begin{equation}\nonumber
\vekdd(k) \,,\; \vekdv(t) \,,\; \vekdw(t) \,,\; \vekx_{\dd i(k)}(t) \,,\; \veku_{\dd i(k)}(t) \,,\; \vekz_{\dd i(k)}(t) \,. 
\end{equation}

Discrete--state switching is bi--uniquely related to {\it discrete--state transitions} \\[-1.5ex]
\begin{equation}\nonumber
\de: \dd_i \rightarrow \dd'_i \,,
\end{equation}
with $\dd(\de)$ the {\it head} of $\de$ and $\dd'(\de)$ the {\it tail} of $\de$. To each $\de$, a set of switching rules ${\mathcal G}^{\de}$ is bi--uniquely assigned. If all rules ${\mathcal G}^{\de}$ are fulfilled, ${\mathcal G}^{\de}$ and, hence, $\de$ becomes active. If, in that case, $\dd(\de)$ is the actually active discrete state, then the discrete--state switching $\dd(\de) \rightarrow \dd'(\de)$ is taking place.  \\[-4ex]

Combined discrete and continuous switching rules are common in hybrid systems. In the present contribution, the switching rules of a transition $\de$ are considered as combined sets of ${\mathcal G}^{\de}_d = {\mathcal G}^{\de}_d(\vekdv)$ involving the discrete input $\vekdv$ and ${\mathcal G}^{\de}_c = {\mathcal G}^{\de}_c(\vekz)$ involving the output $\vekz$ of the continuous subsystem\footnote{${\mathcal G}^{\de}_c$ may also involve time derivatives $\dot \vekz$, $\ddot \vekz$, $...\,$ .}:  \\[-1.5ex]
\begin{equation}\nonumber
{\mathcal G}^{\de}(\vekdv,\vekz) = \{{\mathcal G}^{\de}_d(\vekdv), {\mathcal G}^{\de}_c(\vekz)\} \,.
\end{equation}
Therefore, since a discrete--state switching can depend on $\vekdv$ as well as on $\vekz$, the continuous variable $\vekz$ is interpreted as a further input to the discrete subsystem $\rm A$.

\subsection{The concept of the flat hybrid automaton}
\label{ss:fha_conc}

The class of hybrid automata ({\rm HA}) considered in this contribution is supposed to exhibit the following characteristics: The $\rm HA$ has deterministic dynamical behaviour and all its discrete states $\dd_i$ are reachable. The continuous subsystems are differentially flat \cite{FlLeMaRo1992,FlLeMaRo1995} and the continuous switching rules are deterministic and defined on the flat output $\vekz$. The switching rules are invertible in the sense that, given $\de$, the corresponding activating values of $\vekdv$ and $\vekz$ can be determined explicitly. Thereby, given a sequence of transitions $\de$, explicit determination of the input trajectories becomes possible. This system is called {\it Flat Hybrid Automaton}, $\rm FHA = \{A^{\it fl},C^{\it fl}\}$, combining a {\it flat discrete subsystem} ${\rm A}^{fl}$ and a {\it flat continuous subsystem} ${\rm C}^{fl}$. In the following sections, the $\rm FHA$ is successively derived based on the concept of hybrid automata, introduced in \cite{AlCoHaHeHoNiOlSiYo1995,AlCoHeHo93} and further elaborated in e.g., \cite{HedeSchuLuLa2010}.

\subsection{Continuous subsystem}
\label{ss:fha_c}

The continuous subsystem ${\rm C}$ of a $\rm HA$ is the set of all continuous subsystems ${\rm C}_{\dd i}$ each of which is bi--uniquely assigned to a discrete state $\dd_i$: ${\rm C} = \{ {\rm C}_{\dd 1}, {\rm C}_{\dd 2}, \, ... \, , {\rm C}_{\dd{n\dd}} \}$. Each ${\rm C}_{\dd i}$ represents a 5-tuple ${{\rm C}_{\dd i}} = \{\mathcal{X}_{\dd i},\mathcal{U}_{\dd i},\mathcal{Z}_{\dd i},\mu_{\rm C}, \vekx_0\}$, the elements of which are: 

$\mathcal{X}_{\dd i}$ the continuous state space with $\mathcal{X}_{\dd i} \subseteq {\mathbb R} ^{nx_{\dd i}}$, $nx_{\dd i} \in \mathbb{N}_0$, which is bi--uniquely assigned to the discrete state $\dd_i \in D$, with

\begin{itemize}

\item $\vekx_{\dd i} \in \mathcal{X}_{\dd i}$, ${\rm dim}(\vekx_{\dd i}) = nx_{\dd i}$, the vector of continuous states, whereat $\vekx = \vekx_{\vekdd}(t) := \left( \vekx_{\dd i}| {\dd_i(k) = 1} \right)$ denotes the actually active vector of continuous states, 

\item $\mathcal{X}_{D} = \{ \mathcal{X}_{\dd 1}, \mathcal{X}_{\dd 2}, \, ... \, , \mathcal{X}_{\dd{n\dd}} \}$, the set of all continuous state spaces $\mathcal{X}_{\dd i}$, 

\end{itemize}

$\mathcal{U}_{\dd i}$ the continuous input space $\mathcal{U}_{\dd i} \subseteq {\mathbb R} ^{nu_{\dd i}}$, $nu_{\dd i} \in \mathbb{N}^+$, which is bi--uniquely assigned to a discrete state $\dd_i \in D$, with

\begin{itemize}

\item $\veku_{\dd i} \in \mathcal{U}_{\dd i}$, ${\rm dim}(\veku_{\dd i}) = nu_{\dd i}$, the vector of continuous inputs, whereat $\veku = \veku_{\vekdd}(t) := \left( \veku_{\dd i}| {\dd_i(k) = 1} \right) $ denotes the actually active vector of continuous inputs, and 

\item $\mathcal{U}_{D} = \{ \mathcal{U}_{\dd 1}, \mathcal{U}_{\dd 2}, \, ... \, , \mathcal{U}_{\dd{n\dd}} \}$, the set of all continuous input spaces $\mathcal{U}_{\dd i}$, and 

\end{itemize}

$\mathcal{Z}_{\dd i}$ the continuous output space $\mathcal{Z}_{\dd i} \subseteq {\mathbb R} ^{nz_{\dd i}}$, $nz_{\dd i} \in \mathbb{N}^+$, which is bi--uniquely assigned to a discrete state $\dd_i \in D$, with

\begin{itemize}

\item $\vekz_{\dd i} \in \mathcal{Z}_{\dd i}$, ${\rm dim}(\vekz_{\dd i}) = nz_{\dd i}$, the vector of continuous outputs, whereat $\vekz = \vekz_{\vekdd}(t) := \left( \vekz_{\dd i}| {\dd_i(k) = 1} \right) $ denotes the actually active vector of continuous outputs, 

\item $\mathcal{Z}_{D} = \{ \mathcal{Z}_{\dd 1}, \mathcal{Z}_{\dd 2}, \, ... \, , \mathcal{Z}_{\dd{n\dd}} \}$, the set of all continuous output spaces $\mathcal{Z}_{\dd i}$, and  

\item $\mathcal{Z}^{\rm inv}_{\dd i} \subseteq \mathcal{Z}_{\dd i}$, the {\it continuous invariant output space}, for which $\vekz_{\dd i}$ does not fulfill any set of continuous switching rules $\mathcal G^{\de}_c$: $\mathcal{Z}^{\rm inv}_{\dd i} = \{ \vekz_{\dd i} | \mathcal G^{\,\dd_i\,\rightarrow\, \dd_i'}_c \neq 1 \}$.

\end{itemize}

$\mu_{\rm C}$ denotes a relation that bi--uniquely assigns, to each $\dd_i$, the vector field $\vekf_{\dd i}: \vekxdot_{\dd i} = \vekf_{\dd i}( \vekx_{\dd i}, \veku_{\dd i} )$ that is well defined for all $\vekx_{\dd i}, \veku_{\dd i}$ such that a unique solution $\vekx_{\dd i}(t)$, $t \in [t_0,t^{\star}]$, exists and is Lipschitz given $\vekx_{\dd i}(t_0)$ and $\veku_{\dd i}(t)$. 

Finally, $\vekx_{0} = \vekx(t_0)$, $\vekx_0 \in \mathcal{X}_{D}$, is the initial continuous state, in correspondence with the initial discrete state $\vekdd_0$.

\subsection{Differential flatness of the continuous subsystem}
\label{ss:fha_c_fl}

The continuous subsystem $\rm C$ is considered to fulfil Property \ref{p:prop_F_Phi_Psi} which is introduced in the following, based on the definition of differential flatness. 

\Property{\rm The state--space model $\vekxdot_{\dd i} = \vekf_{\dd i}( \vekx_{\dd i}, \veku_{\dd i} )$ has a bijective output function \\[-3ex]
\begin{equation}\nonumber
\vekz_{\dd i} = F_{\dd i} \left( \vekx_{\dd i}, \veku_{\dd i}, {\dot \veku}_{\dd i}, {\ddot \veku}_{\dd i}, \, ... \, , {\veku_{\dd i}}^{(a_{\dd i})} \right) \,,
\end{equation}
with $nz_{\dd i} = nu_{\dd i}$. Furthermore, bijective functions $\Phi_{\dd i}$ and $\Psi_{\dd i}$ exist and can explicitly be derived, which establish a unique mapping of the output $\vekz_{\dd i}$ and its time derivatives to the state $\vekx_{\dd i}$ and input $\veku_{\dd i}$, respectively,  \\[-2ex]
\begin{equation}\nonumber
\bega{l}
\vekx_{\dd i} = \Phi_{\dd i} \left( \vekz_{\dd i}, {\dot \vekz}_{\dd i}, {\ddot \vekz}_{\dd i}, \, ... \, , {\vekz_{\dd i}}^{(b_{\dd i})} \right) \;\;\;\;\;\;\;\;\;\;\;\;\;\;\;\; \\ 
\veku_{\dd i} = \Psi_{\dd i} \left( \vekz_{\dd i}, {\dot \vekz}_{\dd i}, {\ddot \vekz}_{\dd i}, \, ... \, , {\vekz_{\dd i}}^{(c_{\dd i})} \right)\,. 
\ea
\end{equation}
The components of $\vekz_{\dd i}$ are differentially independent. 
}{p:prop_F_Phi_Psi}

If Property \ref{p:prop_F_Phi_Psi} is fulfilled, ${\rm C}_{\dd i}$ is said to be {\it differentially flat} and $\vekz_{\dd i}$ is the {\it flat output}. For a given trajectory $\vekz^*_{\dd i}(t)$, $t \in [ t_0, t^{\star} ]$, the continuous--input trajectory $\veku^*_{\dd i}(t)$ and the continuous--state trajectory $\vekx^*_{\dd i}(t)$, $t \in [ t_0, t^{\star} ]$, exist, are unique and can be explicitly calculated from $\Phi$ and $\Psi$, without integrating differential equations. \cite{FlLeMaRo1992,FlLeMaRo1995}

\Definition{\rm The joint set of continuous subsystems ${{\rm C}_{\dd i}}$, $\dd_i \in D$, is called the {\it (differentially) flat continuous subsystem} ${\rm C}^{fl} = \{ {\rm C}_{\dd 1}, {\rm C}_{\dd 1}, ... , {\rm C}_{\dd n\dd} \}$ of a hybrid automaton, if all ${{\rm C}_{\dd i}}$ fulfil Property \ref{p:prop_F_Phi_Psi}.}{d:def_Cd_fl}

\subsection{Discrete subsystem}
\label{ss:dha_a}

The discrete subsystem, a 5-tuple ${\rm A} = \{D,V,W,\mu_{\rm A},\vekdd_0\}$, includes the sets of discrete states, inputs and outputs, the transition function and the initial discrete state. Since it shall be possible to explicitly determine input trajectories $\dv_i(t)$ from system inversion like it is possible for differentially flat continuous systems, the discrete subsystem is designed accordingly as a ``flat'' discrete subsystem ${\rm A}^{fl}$. The elements of the 5-tuple are described in the following. 

$D = \{ \dd_1, \dd_2, \, ... \, , \dd_{n\dd} \}$ is the non-empty finite set of $n\dd$ discrete states $\dd_i$ (which are the vertices of the associated automaton graph, in the following also denoted as $\dd$), $n\dd \in \mathbb{N}^+$, with \\[-3ex] 

\begin{itemize}

\item $\dd_i (k) \in \{0,1\}$: $\dd_i$ is inactive iff $\dd_i = 0$, and active iff $\dd_i = 1$, and \\[-1ex]

\item $\vekdd(k) \in {\{0,1\}}^{n\dd}$, the vector of discrete states $\dd_i(k)$, representing the discrete--state trajectory. \\[-4ex]

\end{itemize}

$V = \{ \dv_1, \dv_2, \, ... \, , \dv_{n\dv} \}$ is the finite set of $n\dv$ discrete inputs $\dv_i$ (also: $\dv$), $n\dv \in \mathbb{N}_0$, with \\[-4ex]

\begin{itemize}

\item $\dv_i (t) \in \{0,1\}$: $\dv_i$ is inactive iff $\dv_i = 0$, and active iff $\dv_i = 1$, and \\[-1ex]

\item $\vekdv(t) \in {\{0,1\}}^{n\dv}$, the vector of discrete inputs $\dv_i(t)$, the discrete--input trajectory. \footnote{In order to represent a temporally unique sequence of discrete inputs that all successively occur at the same time point $t$, one can use the respective time indications $t',t'',\,...\,$ .} \\[-3ex]

\end{itemize}

$W = \{ \dw_1, \dw_2, \, ... \, , \dw_{n\dw} \}$ is the non--empty finite set of $n\dw$ discrete outputs $\dw_i$ (also: $\dw$), $n\dw \in \mathbb{N}^+$, with \\[-4ex]

\begin{itemize}

\item $\dw_i (t) \in \{0,1\}$: $\dw_i$ is inactive iff $\dw_i = 0$, and active iff $\dw_i = 1$, and \\[-1ex]

\item $\vekdw(t) \in {\{0,1\}}^{n\dw}$, the vector of discrete outputs $\dw_i(t)$, the discrete--output trajectory. \\[-3ex]

\end{itemize}

The outputs are defined by the bijective output function $H_{\rm A}\!:\,(\dd,\de) \, \mapsto \, \dw$, with $\dd$ the head of $\de$ and $n\dw = n\de$: \\[-1ex]
\be
\bega{ll}
\dw(t) = H_{\rm A}\!\left(\dd(\de(t)), \de(t) \right): & {}\\
\;\;\;\;\;\;\;\;\;\; {\rm if}\;\; \de(t)=1 \wedge \dd(\de(t))=1\,, & {\rm then}\;\; H_{\rm A} = 1 \\
\;\;\; {\rm else}\;{\rm if}\;\; \de(t)=0 \vee \dd(\de(t))=0\,, & {\rm then}\;\; H_{\rm A} = 0 \,. \\
\ea
\label{eq:H_A}
\ee \\[-0.5ex]
This output function implies that, at times $t^{\star}=t',t'',...$ of a state switching, $\dw_i(t^\star)=\de_i(t^\star)=1$, otherwise at times $t \neq t',t'',...\,$ , $\dw_i(t)=0$ ($i\in [1,n\de]$), and, furthermore, that only one $\dw_i(t)$ can be active at a time.

$\mu_{\rm A}$, the transition function of $\rm A$, defines the state succession by uniquely assigning a successor state pair $(\vekdd',\vekx')$ (with $\dd' \in D$, $\vekx' \in \mathcal{X}_{\dd'}$) to the actual state $\vekdd$ and $\vekx$ and associated switching rule sets ${\mathcal G}$ (with $\dd \in D$, $\vekx \in \mathcal{X}_{\dd}$): \\[-1ex]
\begin{equation}\nonumber
(\vekdd',\vekx') = \mu_{\rm A} (\vekdd,\vekx,{\mathcal G}) \, . 
\end{equation}
Thereby, the transition function $\mu_{\rm A}$ unites the following elements $E$, $\delta$, $\mathcal{G}$ and $L$: \\[-4ex]

\begin{itemize}

\item $E = \{ \de_1, \de_2, \, ... \, , \de_{n\de} \}$, the non--empty finite set of $n\de$ discrete--state transitions $\de_i \in \{ 0, 1\}$ (also: $\de$), with $n\de \in \mathbb{N}^+$ and $i \in \{1,2,\,...\,,n\de\}$, where $\de = \de(t)$ represent the directed edges of the automaton graph $(D,E)$ of $\rm A$, and which can be active ($\de = 1$) or inactive ($\de = 0$), with $\vekde(t) \in {\{0,1\}}^{n\de}$, the vector of discrete-- state transitions $\de_i(t)$, \\[-1ex]

\item $\delta(\dd, \dd')\!: (\dd, \dd') \mapsto \de $, $\delta = \{ \, \de_i \, | \, \dd(\de_i) = \dd \wedge \dd'(\de_i) = \dd' \, \}$, the incidence function which is called well--posed in the sense that it assigns a pair of discrete states $(\dd,\dd')$ to a set of transitions $\de_i$ that have the same head $\dd(\de_i) = \dd$ and the same tail $\dd'(\de_i) = \dd'$, and prioritises, according to which of those transitions $\de_i$ the discrete--state switching will occur in the case that more than one $\de_i$ are activated simultaneously, where $m(\dd,\dd')$ is the number of transitions which exist between $\dd,\dd' \in D$, \\[-1ex]

\item $\mathcal{G}^{\de}: (\vekdv,\vekz) \mapsto \de$, sets of switching rules, which each are bi--uniquely assigned to a $\de \in E$ and which activate or deactivate $\de$ in the sense that, iff all rules in a $\mathcal{G}^{\de}$ are fulfilled (which is denoted by $\mathcal{G}^{\de} = 1$), then $\de$ becomes active and otherwise, is inactive ($\mathcal{G}^{\de} = 0$). \\[-1ex]

A discrete--state switching $\dd \rightarrow \dd'$ takes place iff $\dd$ is active and at least one of the associated transitions $\de_i \in \delta(\dd, \dd')$ is active. \\[-1ex]

The subset of the rules in $\mathcal{G}^{\de}(\vekdv,\vekz)$ that involve the discrete input $\vekdv$ is called {\it discrete switching rule set} of $\de$, $\mathcal G^{\de}_d = \mathcal G^{\de}_d(\vekdv) \subseteq \mathcal G^{\de}$, with $V_{\de} = \{\, \dv_{i,\de}=1 \,|\, \mathcal G^{\de}_d = 1 \, \}$ the {\it set of switching discrete inputs $\dv_{i,\de}$} of $\de$.  \\[-1ex] 

The subset of rules in $\mathcal{G}^{\de}(\vekdv,\vekz)$ that involve the continuous flat outputs $\vekz = \vekz_{\dd,\de}$ is called {\it continuous switching rule set} of $\de$, $\mathcal G^{\de}_c = \mathcal G^{\de}_c(\vekz) \subseteq \mathcal G^{\de}$, with ${\mathcal Z}_{\dd,\de} = \{\, \vekz_{\dd,\de} \in {\mathcal Z}_{\dd} \,|\, \mathcal G^{\de}_c = 1 \,\}$ the {\it set} of {\it switching continuous flat outputs $\vekz_{\dd,\de}$} of $\de$.\footnote{Since $\vekz$ is in general not unique, a distinct flat output has to be chosen for the construction of the flat continuous subsystem in order to obtain determinism.} \\[-1ex]

$V^{\rm inv}_{\dd_i} = \{ \dv_j=1 \,|\, \mathcal{G}^{\de}_d = 0 \}$, $\dd(\de) = \dd_i$, is called {\it discrete invariant of $\dd_i$}, the set of discrete inputs that do not influence the activation of $\de$. \\[-1ex]

${\mathcal G}^{E} = \cup\, {\mathcal G}^{\de i}$ is the {\it joint set of switching rules of $\rm A$}, with $\de_i\in E$. \footnote{$\delta$, $\mathcal G^{\de}_d$ and $\mathcal G^{\de}_c$ can be represented in form of look--up tables.} \\[-2ex]

{\footnotesize {\bf Remark.} Since switching rules which involve the flat output $\vekz$ will limit the reachability of the continuous state--space, existence analysis for continuous trajectories \cite{FaHaFi2011,FaHaFi2014} may become relevant.} \\[-1ex]

\item $L_{\dw}\!: \vekx' = L_{\dw}(\vekx)$, the continuous--state transition function which, for each $\dw$ of the discrete transitions $\de$, uniquely assigns a continuous--state successor $\vekx' \in \mathcal{X}_{\dd'(\de)}$ to its predecessor $\vekx \in \mathcal{X}_{\dd(\de)}$: For the actually active $\vekx_{\vekdd}(t)$, $L_{\dw}(\vekx_{\vekdd}(t)) := \vekx'_{\vekdd}(t')$ iff the state switching occurs ($\dw(t)=1$), else, $L_{\dw}(\vekx_{\vekdd}(t)) := \vekx_{\vekdd}(t)$ i.e., for $\dw(t)=0$. \\[-2ex]

{\footnotesize {\bf Remark.} The concept of combining state switching and switching rules involving switching continuous flat outputs implies that a continuous-state switching $\vekx_{\dd i} \rightarrow \vekx'_{\dd i'}$ has to show a correspondence in the continuous flat outputs by $\vekx_{\dd i} = \Phi_{\dd i} \left( \vekz_{\dd i}, {\dot \vekz}_{\dd i}, {\ddot \vekz}_{\dd i}, \, ... \, , {\vekz_{\dd i}}^{(b_{\dd i})} \right)$ and $\vekx'_{\dd i'} = \Phi_{\dd i'} \left( \vekz'_{\dd i'}, {\dot \vekz'}_{\dd i'}, {\ddot \vekz'}_{\dd i'}, \, ... \, , {\vekz'_{\dd i'}}^{(b_{\dd i})} \right)$. }

\end{itemize}

$\vekdd_0$ is the initial state $\vekdd_0 = \vekdd(k_0)$, with $\dd_i(k_0) \in D$. \\[-4ex]

\subsection{Paths, sequences and adjacency matrices}
\label{ss:fha_a_pth_sqnc_adjmtrx}

The concepts of paths, sequences and adjacency matrices of automata and discrete systems are useful to handle reachability analysis and explicit determination of input trajectories by system inversion. The concepts are, therefore, described in the following and are related to trajectory planning in the subsequent sections. \\[-3ex]

A succession of $np$ transitions \\[-1ex]
\begin{equation}\nonumber
P = \{ \de_{\varsigma 1},\de_{\varsigma 2}, ... , \de_{\varsigma nP} \}\,, \; \de \in E 
\end{equation}
is called {\it path} iff head and tail $\dd_{j},\dd_{j}'$ of each of its transitions are pairwise different and the head of $\de_{\varsigma i+1}$ is the tail of $\de_{\varsigma i}$: $\dd'(\de_{\varsigma i}) = \dd(\de_{\varsigma i+1})$. $P = P_{\dd(\de_{\varsigma 1}),\dd'(\de_{\varsigma nP})}$ is referred to as {\it connecting path} of the {\it starting point} $\dd(\de_{\varsigma 1})$ and the {\it end point} $\dd'(\de_{\varsigma nP})$. The sequence of switching rules along a path $P$ is given by \\[-1ex]
\begin{equation}\nonumber
{\mathcal G}^P = \{ {\mathcal G}^{\de,{\varsigma 1}},{\mathcal G}^{\de,\varsigma 2}, ... , {\mathcal G}^{\de,\varsigma nP} \}\,, \; \de_{\varsigma i} \in P \,. \\[1ex]
\end{equation}
For each ${\mathcal G}^{\de,\varsigma i} \in {\mathcal G}^{P}$, the sets of switching inputs $V_{\de}$ and sets of switching flat continuous outputs ${\mathcal Z}_{\dd, \de}$ are given along $P$ through ${\mathcal G}^P$ and can be grouped  into the {\it sequence of switching discrete input sets $V_{P}$ and switching continuous flat output sets ${\mathcal Z}_P$} of $P$: \\[-2ex] 
\begin{equation}\nonumber
\bega{l}
\;\;\; (V_{P},{\mathcal Z}_P) = \\
\;\;\;\;\;\;\; \{ (V_{\de_{\varsigma 1}},{\mathcal Z}_{\dd, \de_{\varsigma 1}}),(V_{\de_{\varsigma 2}},{\mathcal Z}_{\dd, \de_{\varsigma 2}}), ... , (V_{\de_{\varsigma nP}},{\mathcal Z}_{\dd, \de_{\varsigma nP}}) \} \,, 
\ea
\end{equation}
with $\de_{\varsigma i} \in P$. Hence, $(V_{P},{\mathcal Z}_P)$ represents the inputs of the discrete subsystem $\rm A$ in form of the sequence of switching discrete inputs $\dv_{i,\de_{\varsigma i}}$ and switching continuous flat outputs $\vekz_{\dd_j,\de_{\varsigma i}}$, the successive control of which activates the successions of transitions $\de_{\varsigma i}$ of $P$. 

The succession of discrete states \\[-1ex]
\begin{equation}\nonumber
S = S_{\dd_{\xi 1}, \dd_{\xi nS}} = \{ \dd_{\xi 1}, ... , \dd_{\xi nS} \} \,,\, \dd_{\xi i} \in D \\[1ex]
\end{equation}
is called a {\it discrete--state sequence} that is feasible for $\rm A$ iff at least one connecting path $P_{\dd_{\xi 1}, \dd_{\xi nS}}$ exists.  

The time--invariant {\it adjacency matrix} $\matA = \matA({\rm A})$, with ${\rm dim}(\matA) = (n\dd,n\dd)$, is determined by: \\[-1ex]
\begin{equation}\nonumber
\matA = (a_{i,j}) = 
\left\{ 
\bega{cl}
m(\dd_i,\dd_j) & {\rm if} \;\, \exists \, \de \in E: \delta(\dd_i,\dd_j) = \de \\
0 & {\rm if} \not\exists \, \de \in E: \delta(\dd_i,\dd_j) = \de \;\;\; ,
\ea
\right. 
\end{equation}
with $m(\dd_i,\dd_j)$ according to Section \ref{ss:dha_a}.

\subsection{Flat discrete subsystem}
\label{ss:fha_a_fl}

In accordance with differential flatness of the continuous subsystem, it shall be possible to determine of $\rm A$ the input and discrete--state trajectories, i.e., $(V_{P},{\mathcal Z}_P)$ and $S$, from a given output trajectory $\vekdw(t)$, based on system inversion. Furthermore, according to Section \ref{ss:fha_conc}, $D$ shall be reachable and the dynamical behaviour of $\rm A$ shall be deterministic (cf. Section \ref{ss:hybrd_dyn}). In this subsection, the respective system properties, explicit trajectory planning and the definition of the flat discrete subsystem are developed.  

{\bf Reachability.} A discrete subsystem ${\rm A} = \{D,V,\mu_{\rm A},\vekdd_0\}$ as described in Section \ref{ss:dha_a}, is considered reachable if the following property holds:  \\[-2ex]

\Property{\rm From any initial state $\dd_0 \in D$, every other state $\dd \in D$ is reachable in the sense that, for all pairs $\dd_{i1}, \dd_{i2} \in D$, there exists at least one path $P$ that connects $\dd_{i1}$ and $\dd_{i2}$ along the sequence $S(P)$.}{p:prop_reachblt_A}

Reachability according to Property \ref{p:prop_reachblt_A} is given, if the automaton graph of $\rm A$ is strongly connected, which is the case iff the adjacency matrix $\matA({\rm A})$ is irreducible \cite{BrePle1979}. To verify irreducibility, the following criterion given in \cite{Me2000} can be applied: If $(\matI + \matA)^{(n\dd-1)} > {\Mbf 0}$ holds, then $\matA$ (with $a_{i,j} \geq 0$) is irreducible \cite{Ga1959} and, hence, $\rm A$ is reachable, since it is strongly connected.

{\bf Deterministic dynamical behaviour.} The dynamical behaviour of the discrete subsystem $\rm A$ is determined by its transition function $\mu_{\rm A}$. Since $\mathcal G$ in $\mu_{\rm A}$ involves $\vekdv$ and $\vekz$, these latter two variables act as inputs to the discrete subsystem. \\[-2ex]

\Property{\rm The discrete subsystem $\rm A$ of a hybrid automaton $\rm HA$, designed according to Section \ref{ss:dha_a}, shows deterministic dynamical behaviour in the sense that the trajectory of discrete states $\vekdd(k)$ is uniquely determined by $\vekdv(t)$ and $\vekz(t)$, given the initial state $\vekdd_0$.}{p:dtrmn_A}

Property \ref{p:dtrmn_A} holds since, according to Section \ref{ss:dha_a}, all feasible discrete--state transitions $\de$ as well as their prioritisation are uniquely determined by $\delta$ and since, by ${\mathcal G}^E$, a set of deterministic switching rules is by--uniquely assigned to each $\de$. Thereby, it is uniquely prescribed when $\de$ becomes active by respective $\vekdv$ and $\vekz$ such that a discrete--state switching becomes possible. The continuous successor states are uniquely defined by $L_{\dw}$.

{\bf Explicit trajectory planning.} In order to explicitly determine state and input trajectories $S(P)$ and $(V_P,{\mathcal Z}_P)$ from an output trajectory $\vekdw$, a respective inversion of $\rm A$ is proposed. \\[-5ex]

Consider $\rm A$ designed according to Section \ref{ss:dha_a}. The output function $\dw = H_{\rm A}(\dd,\de)$ implies that, at switching times $t^{\star}$, the path $P$ corresponds to the respective succession of discrete outputs $\dw_i(t^\star)$. Hence, for a given output trajectory at times $t^{\star}=t',t'',...$ the trajectory of discrete transitions directly follows by \\[-1ex]
\begin{equation}\nonumber
\vekdw(t'), \vekdw(t''), ... = \vekde(t')|_{\dd_i(\de_j)=1}, \vekde(t'')|_{\dd_i(\de_j)=1}, \, ... \; \,.\\[0ex]
\end{equation} 

Consider $\delta$ and $\mathcal{G}^{\de_i}$ invertible in the sense that \\[-1ex]
\be
\bega{rl}
{\delta}^{-1} \, : & \; \de_i \, \mapsto \, (\dd',\dd'') \\ 
(\mathcal{G}^{\de_i})^{-1} \, : & \; \de_i \, \mapsto \, (\vekdv,\vekz)  \,,
\ea
\label{eq:delta_Gei_inv}
\ee  \\[-0.5ex]
which is the case if $\rm A$ is designed according to Section \ref{ss:dha_a}. Then, given a path $P=P_{\dd_{\xi 1},\dd_{\xi nS}}$ and, thereby, the sequence of switching rules ${\mathcal G}^P$, the corresponding discrete--state sequence $S_{\dd_{\xi 1},\dd_{\xi nS}}$ and sequence of switching inputs $(V_{P},{\mathcal Z}_P)$ is given by Equation (\ref{eq:delta_Gei_inv}). Thus, by $(V_{P} ,{\mathcal Z}_{P})$, the sequence of control input variables $\vekdv(t^\star)$ and $\vekz(t^\star)$ at switching times $t^\star = t', t'', \, ... \, $ to realise $S_{\dd_{\xi 1}, \dd_{\xi nS}}$ is uniquely determined. 

\vspace{-4ex}
Hence, in analogy to trajectory planning for differentially flat continuous systems, state and input trajectories of the discrete subsystem  \\[-1ex]
\begin{equation}\nonumber
\bega{c}
\vekdd(k_1), \vekdd(k_2), ... \\
(\vekdv(t'),\vekz(t')), (\vekdv(t''),\vekz(t'')), ... \\
\ea
\end{equation} 
are determined from a given output trajectory $\vekdw(t')$, $\vekdw(t'')$, $...$ through the following steps: \\[-4ex]

Given $\vekdw(t'), \vekdw(t''), \, ... = \vekde(t'), \vekde(t''), \, ...$ then, \\[-4ex] 

\begin{itemize}

\item the associated discrete--state trajectory is determined straight--forwardly through the inversion of the incidence function $\delta^{-1}$: \\[-2ex] 
\begin{equation}\nonumber
\bega{l}
\vekdw(t'), \vekdw(t''), ... \\
\;\;\;\;\;\;\;\; \Rightarrow \vekdd(k_1), \vekdd(k_2), ... \,, \\
\ea
\end{equation} 

\item from the inverse of the discrete switching rule sets, the associated sequence of sets of switching discrete inputs is determined, from which the sequence of discrete inputs is directly derived: \\[-2ex]
\begin{equation}\nonumber
\bega{l}
\vekdw(t'), \vekdw(t''), ... \\
\;\;\;\;\;\;\;\; \Rightarrow  ({\mathcal G}^{\vekdw(t')}_d)^{-1}, ({\mathcal G}^{\vekdw(t'')}_d)^{-1}, ... \\
\;\;\;\;\;\;\;\; \Rightarrow  V_{\vekdw(t')}, V_{\vekdw(t'')}, ... \\ 
\;\;\;\;\;\;\;\; \Rightarrow  \vekdv(t'), \vekdv(t''), ...  \,, 
\ea
\end{equation} 

\item from the inverse of the continuous switching rule sets, the associated sequence of sets of switching continuous flat outputs are determined, from which each a value of the respective switching flat continuous outputs is determined:  \\[-2ex]
\begin{equation}\nonumber
\bega{l}
\vekdw(t'), \vekdw(t''), ... \\
\;\;\;\;\;\;\;\; \Rightarrow  ({\mathcal G}^{\vekdw(t')}_c)^{-1}, ({\mathcal G}^{\vekdw(t'')}_c)^{-1}, ... \\
\;\;\;\;\;\;\;\; \Rightarrow  {\mathcal Z}_{d(\vekdw(t')),\vekdw(t')}, {\mathcal Z}_{d(\vekdw(t'')),\vekdw(t'')}, ...  \\ 
\;\;\;\;\;\;\;\; \Rightarrow  \vekdz(t'), \vekdz(t''), ...  \,. 
\ea
\end{equation} 

\end{itemize}

Subsystem $\rm A$ designed as above exhibits Property \ref{p:inpt_traj_sys_inv}.  \\[-4ex]

\Property{\rm Since the inverse of the incidence function $\delta$ and the switching rules ${\mathcal G}$ exist and can be explicitly derived for a discrete subsystem $\rm A$ of a $\rm HA$ designed according to Section \ref{ss:dha_a}, the trajectories of $\vekd(k)$ and $(\vekdv(t^\star),\vekz(t^\star))$ can be explicitly determined for any feasible path $P_{\dd_{i1},\dd_{i2}}$ of $\rm A$, if    the discrete output of $\rm A$ is set according to Equation (\ref{eq:H_A}). With this property $\rm A$ is said to be {\it explicitly schedulable}.} {p:inpt_traj_sys_inv} 

{\bf Defining the flat discrete subsystem.} If the properties as described above hold, then $\rm A$ is reachable, deterministic and invertible such that the discrete--state trajectory and the switching inputs can explicitly be determined from the output trajectory of a feasible path of $\rm A$.  

\Definition{\rm If Properties \ref{p:prop_reachblt_A}, \ref{p:dtrmn_A} and \ref{p:inpt_traj_sys_inv} hold, then $\rm A$ is called the {\it flat discrete subsystem} ${\rm A}^{fl}$ of a hybrid automaton. }{d:def_Afl}

\subsection{Definition of the Flat Hybrid Automaton}
\label{ss:fha_dev_remarks}

If a $\rm HA$ consists of a flat continuous and a flat discrete subsystems according to Definition \ref{d:def_Cd_fl} and \ref{d:def_Afl}, then $\vekdv(t)$ and $\veku(t)$ can explicitly be determined if the initial state pair $(\vekdd_0,\vekx_0)$ and a discrete target state $\dd_{final}$ with a target flat output $\vekz_{\dd_{final}}$ are given.

\Definition{\rm If the continuous subsystem of a hybrid automaton $\rm HA$ is a {\it differentially flat continuous subsystem} and the discrete subsystem is a {\it flat discrete subsystem}, then the resulting dynamical system is called {\it flat hybrid automaton} ${\rm FHA}= \{{\rm A}^{fl},{\rm C}^{fl}\}$.}{d:def_fha}


\section{Construction and trajectory planning}
\label{s:constr_trajplan}

The concept of flat continuous and discrete subsystems yields a relatively straight--forward approach for the construction a $\rm FHA$, summarised in the following steps: \\[-4ex]

\begin{itemize}

\item From the physical continuous system model given by a set of differential equations involving the time--derivatives of the continuous state variables, a state--space model is derived including all switching terms. 

\item All possible switching configurations of the state--space model are specified to obtain $\vekf_{\dd i}$, from which the discrete state--space $D$ and input--space $V$ as well as the corresponding continuous subsystems ${\rm C}_{\dd i}$ are derived. The continuous subsystems shall be differentially flat according to Property \ref{p:prop_F_Phi_Psi}, such that Definition \ref{d:def_Cd_fl} is fulfilled. \footnote{Eventually, the respective system to-be-controlled has to be designed such that it is flat (using \cite{WaZe2008,WaZe2010}). The authors argue that this can be the necessary price for taming the complexity of hybrid systems for technical application.}

\item In order to obtain $\mu_{\rm A}$, discrete--state transitions $E$, the incidence function $\delta$, the switching rules ${\mathcal G}^{E}$ and the continuous-state transition function $L_{\dw}$ are derived. The discrete subsystem $\rm A$ shall be well posed in the sense that it is reachable, has deterministic dynamical behaviour and is explicitly schedulable according to Properties \ref{p:prop_reachblt_A}, \ref{p:dtrmn_A} and \ref{p:inpt_traj_sys_inv}, such that $\rm A$ is a flat discrete subsystem according to Definition \ref{d:def_Afl}. \\[-4ex]

\end{itemize}

This construction yields a FHA, the dynamical system of which is represented in the block diagram in Figure \ref{f:bd_fha}. The algorithm for explicit input trajectory determination can be designed as follows: \\[-4ex]

\begfig{8.5cm}
\psfrag{vw}{\scriptsize {$\vekdw$}}
\psfrag{vv}{\scriptsize {$\vekdv$}}
\psfrag{L_e_x_vd}{\scriptsize {$L_{\dw}(\vekx_{\vekd})$}}
\psfrag{vd}{\scriptsize {$\vekdd$}}
\psfrag{x_vd}{\scriptsize {$\vekx_{\vekd}$}}
\psfrag{z_vd}{\scriptsize {$\vekz_{\vekd}$}}
\psfrag{A_fl}{{${\rm A}^{fl}$}}
\psfrag{C_fl}{{${\rm C}^{fl}$}}
\psfrag{u_vd}{\scriptsize {$\veku$}}
\psfrag{x0d=L_e_x_vd}{\scriptsize {$\vekx_{0,\vekdd}=L_{\dw}(\vekx_{\vekdd})$}}
\psfrag{A_dynamics_1}{\scriptsize {$(\vekdd',\vekx') = \mu_{\rm A} (\vekdd,\vekx_{\vekdd},{\mathcal G(\vekdv,\vekz_{\vekdd})})$}}
\psfrag{A_dynamics_2}{\scriptsize {(${\rm with\!}:\,\vekx'_{\vekdd} = L_{\dw}(\vekx_{\vekdd})$)}}
\psfrag{A_dynamics_3}{\scriptsize {$\vekdw = H_{\rm A}(\vekdd,\vekde)$}}
\psfrag{A_dynamics_4}{\scriptsize {$\vekdd_0$}}
\psfrag{C_dynamics_1}{\scriptsize {$\vekxdot_{\vekdd} = \vekf_{\vekdd}(\vekx_{\vekdd}, \veku_{\vekdd})$}}
\psfrag{C_dynamics_2}{\scriptsize {$\vekz_{\vekdd} = F_{\vekdd}\!\left(\vekx_{\vekdd}, \veku_{\vekdd}, {\dot \veku}_{\vekdd}, .. , {\veku_{\vekdd}}^{(a_{\vekdd})}\right)$}}
\psfrag{C_dynamics_3}{\scriptsize {$\vekx_{0,\vekdd}$}}
\efig{7cm}{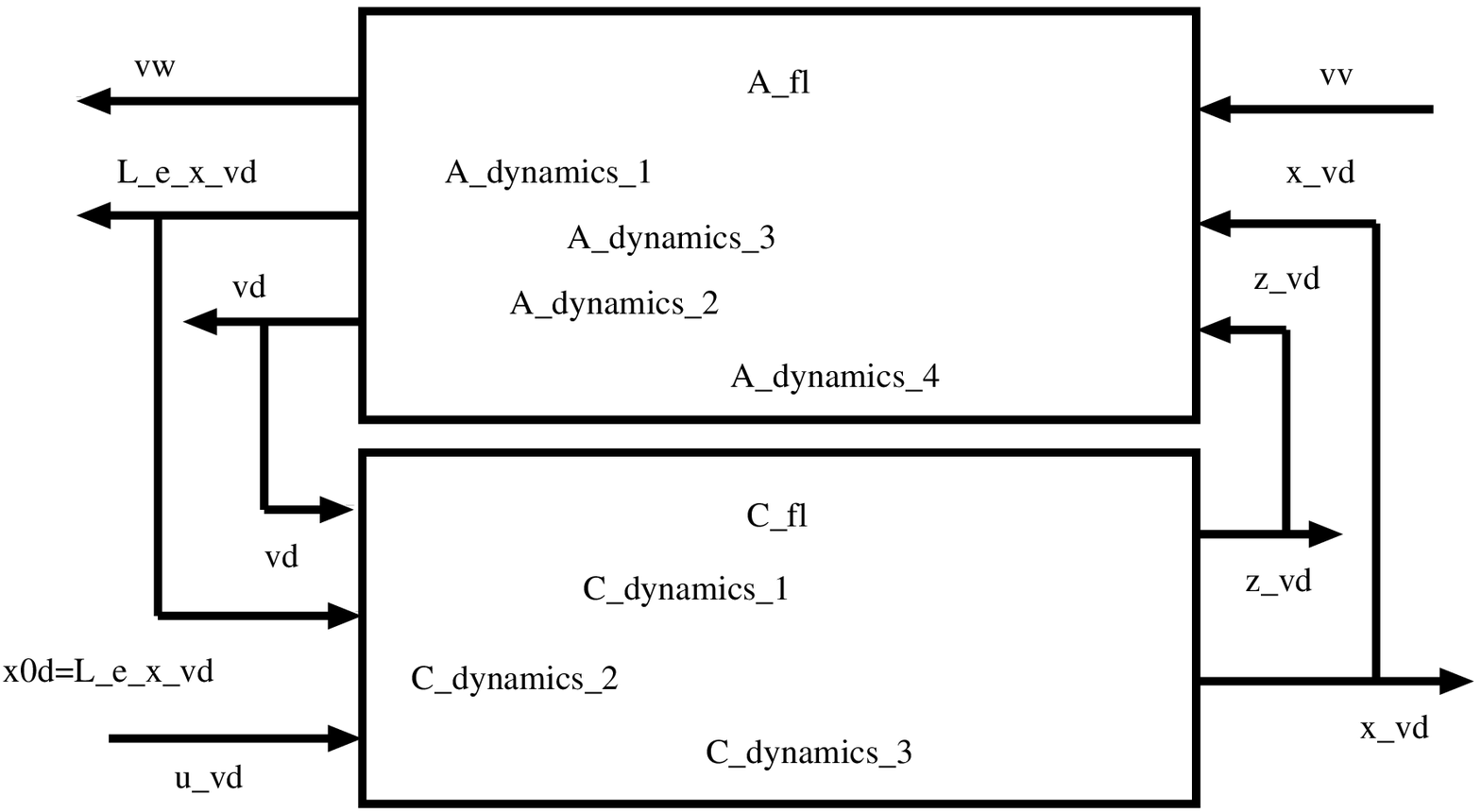}{Block diagram of the FHA with the actually active system variables}{f:bd_fha}

{\bf Algorithm 1} Given a starting point $\dd_0 = \dd_{\xi 1}$ and $\vekx_{0,\dd_0}$ and an end point with $\dd_{final} = \dd_{\xi nS}$ and $\vekz_{final} \in {\mathcal Z}^{\rm inv}_{\dd_{final}}$. Then, \\[-5ex]

\begin{enumerate}

\item determine the paths $P_{\dd_{0},\dd_{final}}$ for feasible discrete--state sequences $S_{\dd_{0},\dd_{final}}$, \\[-1ex]

\item select a discrete--output sequence $\dw_i(t^\star)$ with corresponding $P$ and $S$ from $P_{\dd_{0},\dd_{final}}$ and $S_{\dd_{0},\dd_{final}}$, \\[-1ex]

\item determine the continuous invariant output spaces $\mathcal{Z}^{\rm inv}_{\dd_{\xi i}}$ and, from the sequence of switching rules $\mathcal{G}^{P}$, determine $\mathcal{G}^{\de_{\varsigma i}}_d$ and $\mathcal{G}^{\de_{\varsigma i}}_c$ together with the set of switching continuous outputs $\mathcal{Z}_{\dd_{\xi i},\de_{\varsigma i}}$, for each $\dd_{\xi i} \in S$ and $\de_{\varsigma i} \in P$, \\[-1ex]

\item from $\mathcal{G}^{\de_{\varsigma i}}_d$ and $\mathcal{G}^{\de_{\varsigma i}}_c$, determine the sequence of switching inputs and switching flat outputs $(V_P,{\mathcal Z}_P)$ according to Section \ref{ss:fha_a_pth_sqnc_adjmtrx}, \\[-1ex]

\item for $\dd_0$ ($\dd_0 = \dd_{\xi 1} \in S$), do:

\begin{itemize}

\item determine the initial flat output $\vekz_{0,\dd}$ and the switching flat output $\vekz_{\dd,\de}$ (for $\dd=\dd_0$ and $\de=\de_{\varsigma 1}$) \\[-0.5ex]
\begin{equation}\nonumber
\bega{l}
\vekz_{0,\dd} = F \left( \vekx_{0,\dd}, \veku_{\dd}, {\dot \veku}_{\dd}, {\ddot \veku}_{\dd}, \, \ldots \, , {\veku}^{(a_{\dd})}_{\dd} \right) \in {\mathcal Z}^{\rm inv}_{\dd} \\
\;\;\; {\rm via \; choice \; of\;} \veku_{\dd}, {\dot \veku}_{\dd}, {\ddot \veku}_{\dd}, \, \ldots \, , {\veku}^{(a_{\dd})}_{\dd} \,,\; {\rm and} \\
\vekz_{\dd,\de} \in {\mathcal Z}_{\dd,\de}\; ({\rm for} \;\, {\mathcal Z}_{\dd,\de} \in {\mathcal Z}_{P})\;, \\
\ea
\end{equation} 
\item choose $t^\star=t_{\de_{\varsigma 1}}$ and plan a trajectory $\vekz^*_{\dd_0}(t)$ with starting point $\vekz^*_{\dd_0}(t=0) = \vekz_{0,\dd_0}$ and end point $\vekz^*_{\dd_0}(t=t_{\de_{\varsigma 1}}) = \vekz_{\dd_0,\de_{\varsigma 1}}$, 

\item determine $\vekx^*_{\dd_0}(t)$ and $\veku^*_{\dd_0}(t)$, $t \in [ 0, t_{\de_{\varsigma 1}} ]$, from $\Phi_{\dd_0}$ and $\Psi_{\dd_0}$, 

\item for $t = {t'}_{\de_{\varsigma 1}}$, determine $\vekx' = L_{\dw_{\varsigma 1}}(\vekx_{\dd_0}(t_{\de_{\varsigma 1}}))$ which yields $\vekx_{0,\varsigma 1+1} = \vekx'$. \\[-1ex]

\end{itemize}

\item for the subsequent $\dd_{\xi i} \in S$ and $\de_{\varsigma i} \in P$, with $\xi_1 < \xi_i < \xi_{nS}$, repeat (5) in the sense that: \\[-2ex]

\begin{itemize}

\item with $\dd = \dd_{\xi i}$ and $\de = \de_{\varsigma i}$ determine \\[-0.5ex]
\begin{equation}\nonumber
\bega{l}
\vekz_{0,\dd} = F \left( \vekx_{0,\dd}, \veku_{\dd}, {\dot \veku}_{\dd}, {\ddot \veku}_{\dd}, \, \ldots \, , {\veku}^{(a_{\dd})}_{\dd} \right)  \in {\mathcal Z}^{\rm inv}_{\dd} \\
\;\;\; {\rm via \; choice \; of\;} \veku_{\dd}, {\dot \veku}_{\dd}, {\ddot \veku}_{\dd}, \, \ldots \, , {\veku}^{(a_{\dd})}_{\dd} \,,\; {\rm and}  \\
\vekz_{\dd,\de} \in {\mathcal Z}_{\dd,\de}\; ({\rm for} \;\, {\mathcal Z}_{\dd,\de} \in {\mathcal Z}_{P})\;, \\
\ea
\end{equation} 
\item choose $t^\star=t_{\de_{\varsigma i}}$ and plan a trajectory $\vekz^*_{\dd_{\xi i}}(t)$ with starting point $\vekz^*_{\dd_{\xi i}}(t=0) = \vekz_{0,\dd_{\xi i}}$ and end point $\vekz^*_{\dd_{\xi i}}(t=t_{\de_{\varsigma i}}) = \vekz_{\dd_{\xi i},\de_{\varsigma i}}$, 

\item determine $\vekx^*_{\dd_{\xi i}}(t)$ and $\veku^*_{\dd_{\xi i}}(t)$, $t \in ] 0, t_{\de_{\varsigma i}} ]$, from $\Phi_{\dd_{\xi i}}$ and $\Psi_{\dd_{\xi i}}$, 

\item for $t = {t'}_{\de_{\varsigma i}}$, determine $\vekx' = L_{\dw_{\varsigma i}}(\vekx_{\dd_{\xi i}}(t_{\de_{\varsigma i}}))$ to obtain $\vekx_{0,\xi i+1} = \vekx'$. \\[-1ex]

\end{itemize}

\item for $\dd = \dd_{\xi_{nS}}$ ($\dd_{\xi_{nS}} = \dd_{final} \in S$), do: \\[-2ex]

\begin{itemize}

\item determine \\[-0.5ex]
\begin{equation}\nonumber
\bega{l}
\vekz_{0,\dd} = F \left( \vekx_{0,\dd}, \veku_{\dd}, {\dot \veku}_{\dd}, {\ddot \veku}_{\dd}, \, \ldots \, , {\veku}^{(a_{\dd})}_{\dd} \right) \in {\mathcal Z}^{\rm inv}_{\dd} \\
\;\;\; {\rm via \; choice \; of\;} \veku_{\dd}, {\dot \veku}_{\dd}, {\ddot \veku}_{\dd}, \, \ldots \, , {\veku}^{(a_{\dd})}_{\dd} \\
\ea
\end{equation} 
\item choose $t_{final}$ and plan a trajectory $\vekz^*_{\dd_{\xi nS}}(t)$ with starting point $\vekz^*_{\dd_{\xi nS}}(t=0) = \vekz_{0,\dd_{\xi nS}}$ and end point $\vekz^*_{\dd_{\xi nS}}(t=t_{final}) = \vekz_{final}$, 

\item determine $\vekx^*_{\dd_{\xi nS}}(t)$ and $\veku^*_{\dd_{\xi nS}}(t)$, $t \in ] 0, t_{final} ]$, from $\Phi_{\dd_{\xi nS}}$ and $\Psi_{\dd_{\xi nS}}$, \\[-1ex]

\end{itemize}

\item For $\varsigma_1 \leq \varsigma_i \leq \varsigma_{nP}$, assign to each switching input $\vekdv_{\de_{\varsigma i}} \in V_P$ the corresponding switching time $t_{\de_{\varsigma i}}$:  \\[-0.5ex]
\begin{equation}\nonumber
\bega{l}
\vekdv_{\de_{\varsigma i}} = \vekdv_{\de_{\varsigma i}}(t_{\de_{\varsigma i}}) \,. \;\;\,\Box \\
\ea
\end{equation} 

\end{enumerate}

Steps (5), (6) and (7) of Algorithm 1 yield the continuous--time trajectories of the continuous input $\veku^*_{\dd}(t)$, state $\vekx^*_{\dd}(t)$ and flat output $\vekz^*_{\dd}(t)$ for all $\dd \in S_{{\dd_0},\dd_{final}}$. Step (8) yields, for all $\de \in P_{\dd_0,\dd_{final}}$, the time sequence of discrete inputs $\vekdv_{\de_{\varsigma i}}(t_{\de_{\varsigma i}})$. Hence, steps (1) through (8) provide, given a start point, end point and a choice of $P$ and switching times $t_{\de_{\varsigma i}}$, the trajectories of the variables $\veku$, $\vekx$, $\vekz$ and $\vekdv$ to realise $S$. For each $\dd_{\xi i} \in S$, $\vekz_{0,\dd_{\xi i}}$ has to be determined according to steps (5) through (7) such that the respective initial continuous states are $\vekx_{0,\dd_{\xi i}}$. This inverts the system.


\section{Examples}
\label{s:fha_examples}

\subsection{Preliminary remarks}
\label{ss:exmpl_rmrks}
Two demonstrative examples are presentedj. Both have the same automaton graph which is strongly connected (Figure \ref{f:autgr_FHA_exmpl}). For trajectory planning the following path containing all discrete--state transitions is chosen \\[-9ex]

\be
P=\{\de_{1},\de_{6},\de_{11},\de_{5},\de_{7},\de_{2},\de_{9},\de_{10},\de_{3},\de_{12},\de_{8},\de_{4}\} \,.
\label{eq:xmpl_P}
\ee \\[-9ex]

It yields the discrete--state sequence \\[-9ex]

\be
S=\{\dd_1,\dd_2,\dd_4,\dd_2,\dd_3,\dd_1,\dd_3,\dd_4,\dd_1,\dd_4,\dd_3,\dd_2,\dd_1\} \,.
\label{eq:xmpl_S}
\ee  \\[-9ex]

\subsection{One--tank system with one continuous input}
\label{ss:one_tank_one_u}

The first example is a one--tank system (Figure \ref{f:one_tank_1_ps}) that is inspired by \cite{FlJoSR2005}. The system setup is as follows:   \\[-4ex]

\begin{itemize}

\item $x$: level $l_1 = z = x \geq 0$,
\item continuous control flow $u$ (in or out),
\item permanent outflow $u_{out,1} = c_{out} \, \sqrt{l_1}$,
\item outflow switchable through $\dv_1$ at level $l_1 = 0$, \\ with $u_{out,\dv1} = c_{1}({\dv}_1) \, \, \sqrt{l_1}$,
\item overflow at level $l_1 = l_0$, active if $l_1 > l_0$, \\ with $u_{ovf} = c_{ovf} \, \sqrt{l_1-l_0}$. \\[-1ex]

\end{itemize}

\begfig{8cm}
\psfrag{1}{\small {$\dd_1$}}
\psfrag{2}{\small {$\dd_2$}}
\psfrag{3}{\small {$\dd_3$}}
\psfrag{4}{\small {$\dd_4$}}
\psfrag{e_1}{\small {$\de_{1}$}}
\psfrag{e_2}{\small {$\de_{2}$}}
\psfrag{e_3}{\small {$\de_{3}$}}
\psfrag{e_4}{\small {$\de_{4}$}}
\psfrag{e_5}{\small {$\de_{5}$}}
\psfrag{e_6}{\small {$\de_{6}$}}
\psfrag{e_7}{\small {$\de_{7}$}}
\psfrag{e_8}{\small {$\de_{8}$}}
\psfrag{e_9}{\small {$\de_{9}$}}
\psfrag{e_10}{\small {$\de_{10}$}}
\psfrag{e_11}{\small {$\de_{11}$}}
\psfrag{e_12}{\small {$\de_{12}$}}
\efig{7cm}{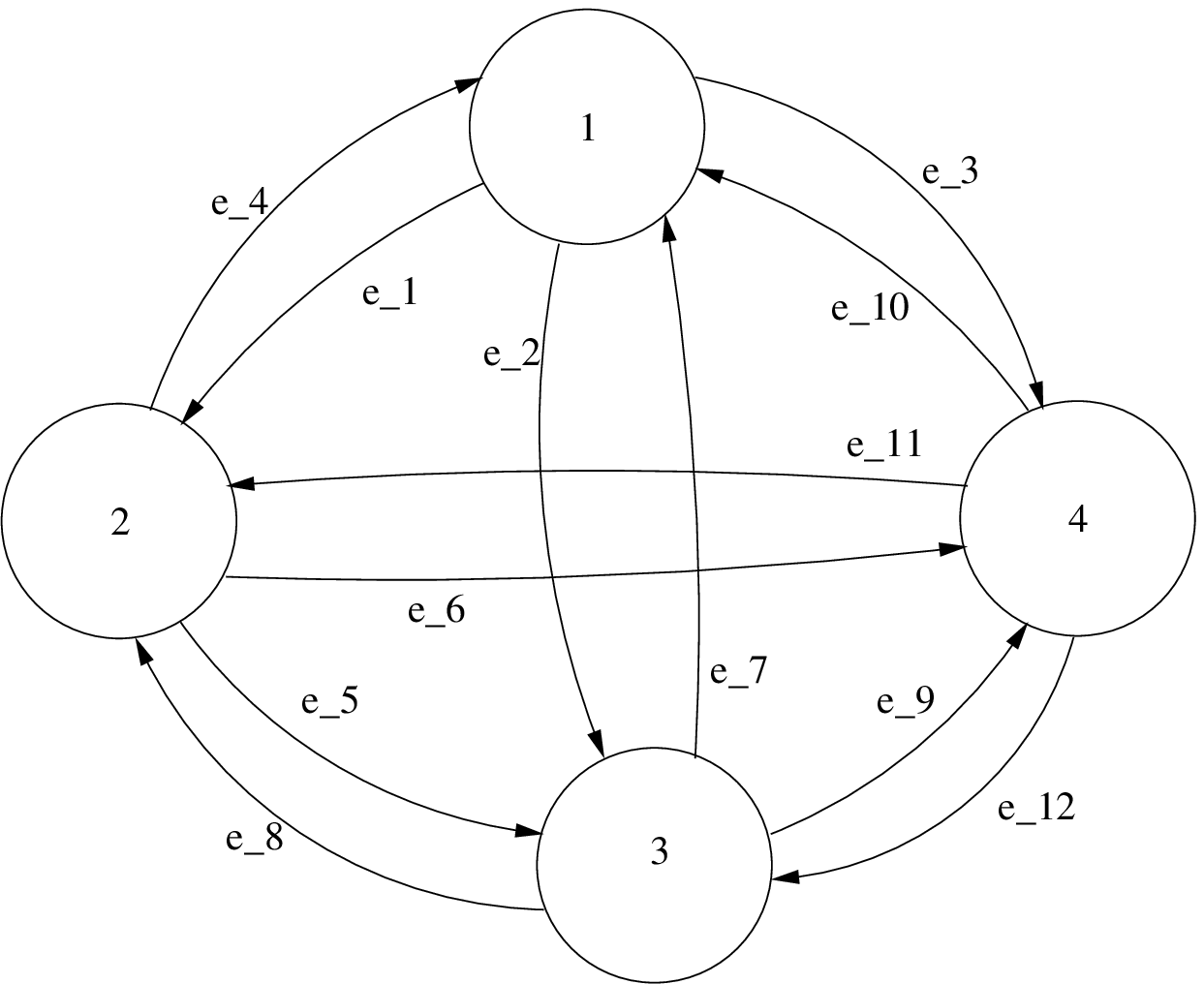}{Automaton graph of the presented examples}{f:autgr_FHA_exmpl}

Equation (\ref{eq:xmpl1_dl1}) represents the dynamical model of the continuous system in $l_1 \in [0,\infty [$, including the switching elements: \\[-1,5ex]
\be
\bega{ccl}
{\dot l}_1 = u & - & c_{out}\, \sqrt{ l_1 } \\ 
{} & - & c_{1}({\dv}_1) \, \, \sqrt{l_1} \\
{} & - & c_{ovf}\, \, H(l_1 - l_0) \, \sqrt{ |l_1 - l_0| } \,,
\ea
\label{eq:xmpl1_dl1}
\ee  \\[-1ex]
with the Heaviside function $H(l_a - l_b)$: \\[-5ex]

\ind{l}{
$H=1$ if $(l_a-l_b)>0$, \\ $H=\frac{1}{2}$ if $(l_a-l_b)=0$, \\
$H=0$ if $(l_a-l_b)<0$, \\[-2ex] 
}

and the outflow switching $c_1(\dv_1)$: \\[-5ex]

\ind{l}{
$c_1(\dv_1) = c_{v,1}$ if ${\dv}_1(t) = 1$, else $c_1(\dv_1) = 0$ \\[-2ex]
}

The initial condition is $l_1(t_0)=l_{1,0}$.\footnote{$[l]=\rm m$, $[u]=\rm \frac{m}{s}$, $[c]=\rm \frac{\sqrt{m}}{s}$}  \\[-4ex]

\begfig{7cm}
\psfrag{Reservoir}{{Reservoir}}
\psfrag{Pump}{{Pump}}
\psfrag{Tank_1}{{Tank $1$}}
\psfrag{u}{{$u$}}
\psfrag{mout_1}{{$u_{out,1}$}}
\psfrag{mout_v1}{{$u_{out,\dv1}$}}
\psfrag{movf}{{$u_{ovf}$}}
\psfrag{l0}{{$l_0$}}
\psfrag{l1}{{$l_1$}}
\psfrag{v1}{{$\dv_1$}}
\efig{7cm}{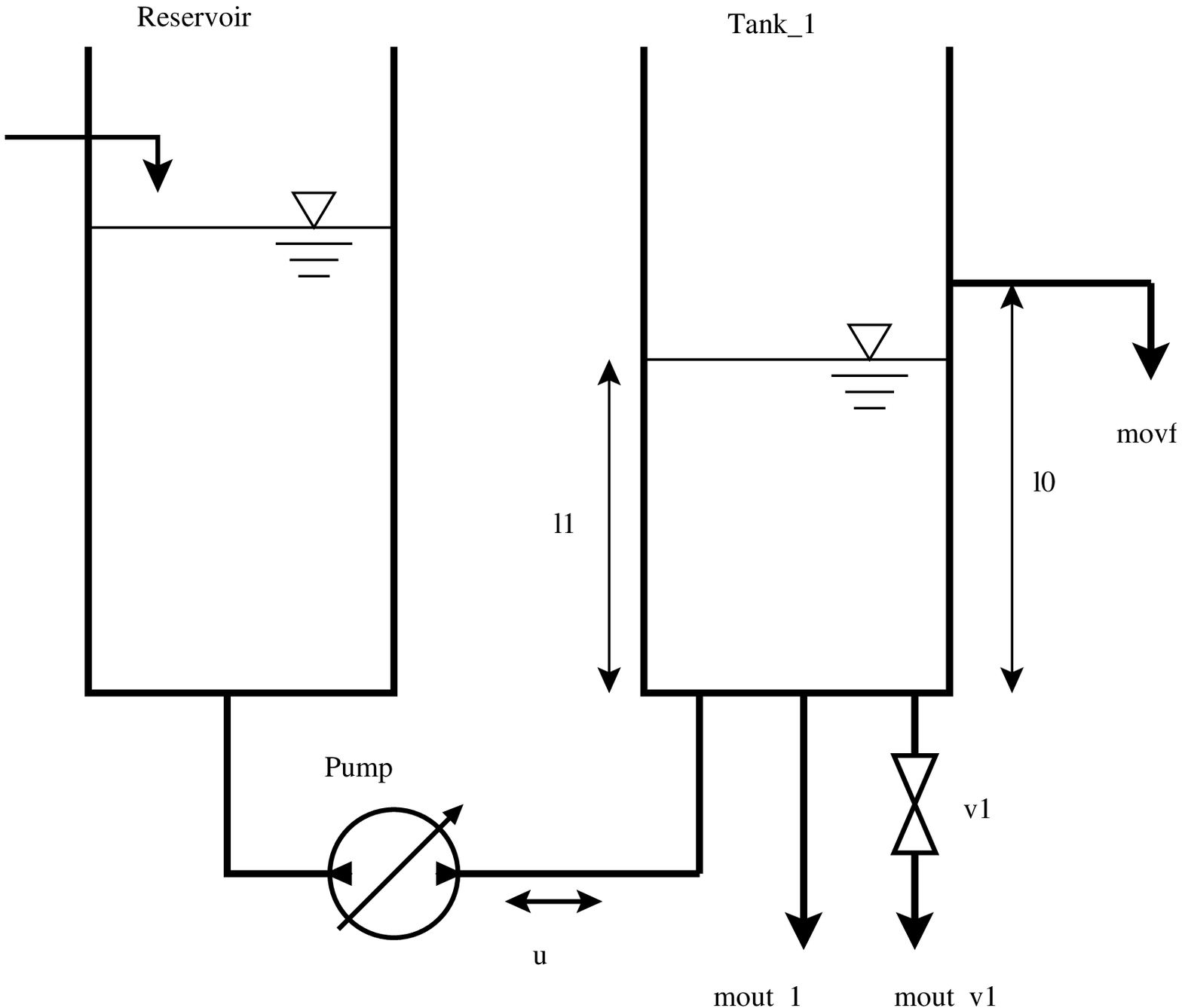}{One--Tank system setup}{f:one_tank_1_ps}

The adjacency list\footnote{An adjacency list is a look--up table that groups all discrete states of an automaton together with their respective successor states and corresponding discrete--state transitions.} is given in Table \ref{tab:xplm1_AL}, extended by the discrete transitions and the respectively associated switching rules as well as $V^{\rm inv}_{\dd_i}$ and $\mathcal{Z}^{\rm inv}_{\dd_i}$ that are valid for the respective discrete states.  \\[-2ex]

\tabcap{htb}{cc|ccl}{
\hline
$\dd_i$ & $\mathcal{Z}^{\rm inv}_{\dd_i}\,,\; V^{\rm inv}_{\dd_i}$ & $\dd'_i$ & $\de_j$ & $\mathcal{G}^{\de_j}$ \\
\hline
$\dd_1$ & $l_1 \leq l_0$ & $\dd_2$ & $\de_1$ & $l_1 \leq l_0\,,\; \dv_1 = 1$ \\[-2ex]
{}      & $\dv_1 = 0$    & $\dd_3$ & $\de_2$ & $l_1 > l_0\,,\; \dv_1 = 0$ \\[-2ex]
{}      & {}             & $\dd_4$ & $\de_3$ & $l_1 > l_0\,,\; \dv_1 = 1$ \\
\hline
$\dd_2$ & $l_1 \leq l_0$ & $\dd_1$ & $\de_4$ & $l_1 \leq l_0\,,\; \dv_1 = 0$ \\[-2ex]
{}      & $\dv_1 = 1$    & $\dd_3$ & $\de_5$ & $l_1 > l_0\,,\; \dv_1 = 0$ \\[-2ex]
{}      & {}             & $\dd_4$ & $\de_6$ & $l_1 > l_0\,,\; \dv_1 = 1$ \\
\hline
$\dd_3$ & $l_1 > l_0$    & $\dd_1$ & $\de_7$ & $l_1 \leq l_0\,,\; \dv_1 = 0$ \\[-2ex]
{}      & $\dv_1 = 0$    & $\dd_2$ & $\de_8$ & $l_1 \leq l_0\,,\; \dv_1 = 1$ \\[-2ex]
{}      & {}             & $\dd_4$ & $\de_9$ & $l_1 > l_0\,,\; \dv_1 = 1$ \\
\hline
$\dd_4$ & $l_1 > l_0$    & $\dd_1$ & $\de_{10}$ & $l_1 \leq l_0\,,\; \dv_1 = 0$ \\[-2ex]
{}      & $\dv_1 = 1$    & $\dd_2$ & $\de_{11}$ & $l_1 \leq l_0\,,\; \dv_1 = 1$ \\[-2ex]
{}      & {}             & $\dd_3$ & $\de_{12}$ & $l_1 > l_0\,,\; \dv_1 = 0$ \\
}{8cm}{Extended adjacency list of the One--Tank}{tab:xplm1_AL}

For the four discrete states, $F_{\dd_i}$ is $l_1=z$ and $\Phi_{\dd_i}$ is $z=l_1$. $\Psi_{\dd_i}$ is derived from the continuous dynamics (Equation (\ref{eq:xmpl1_dl1})), cf. Table \ref{tab:xplm1_di_muc}. \\[-2ex]

\tabcap{ht}{cl}{
\hline
$\dd_i$ & $\Psi_{\dd_i}$ \\
\hline
$\dd_1$ & $u = {\dot l}_1 + c_{out}\,\sqrt{l_1}$ \\[-1.5ex]
$\dd_2$ & $u = {\dot l}_1 + c_{out}\,\sqrt{l_1} + c_{v,1}\,\sqrt{l_1}$ \\[-1.5ex]
$\dd_3$ & $u = {\dot l}_1 + c_{out}\,\sqrt{l_1} + c_{ovf}\,\sqrt{l_1-l_0}$ \\[-1.5ex]
$\dd_4$ & $u = {\dot l}_1 + c_{out}\,\sqrt{l_1} + c_{ovf}\,\sqrt{l_1-l_0}+c_{v,1}\,\sqrt{l_1}$ \\[1ex]
}{9cm}{$\Psi_{\dd_i}$ for the one--tank example}{tab:xplm1_di_muc}

Using e.g., $z(t)= a\cdot (t-t_0) + b$ provides, together with the adjacency list and $\Psi$, the explicit expressions to completely schedule the system trajectories. All required {\rm FHA} properties are fulfilled, hence, the one--tank example is a flat hybrid automaton according Definition \ref{d:def_fha}. \\[-1ex]

\begfig{9cm}
\efig{9cm}{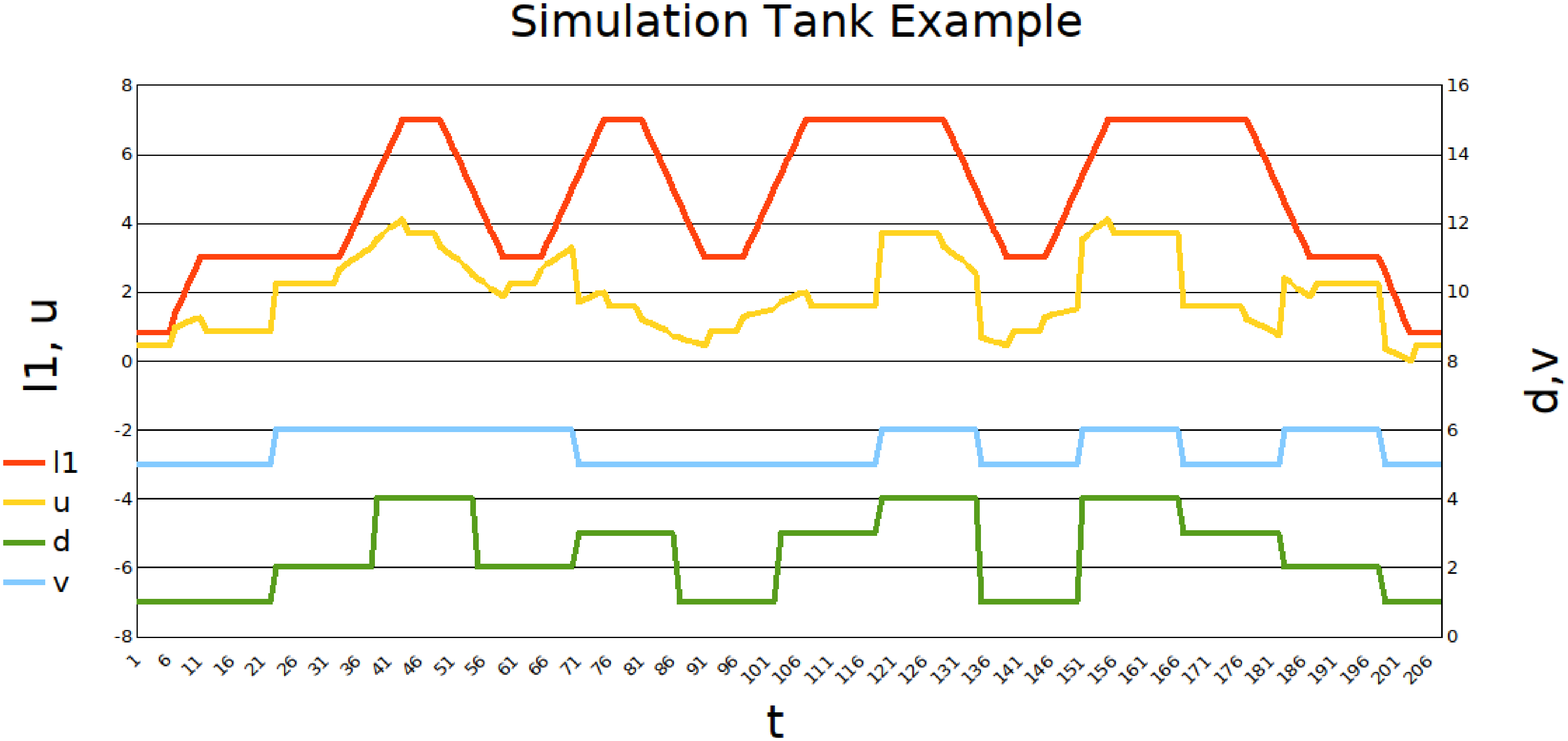}{Simulation of the one--tank example}{f:sim_tank}

Trajectory planning and simulation is set out as follows: For the path $P$ with the feasible sequence $S$ (Equations (\ref{eq:xmpl_S}) and (\ref{eq:xmpl_P})), applying Table \ref{tab:xplm1_AL} provides the sequence of switching rules to realise $P$, yielding $\dv(t^{\star})$ and $\vekz_{\vekdd}(t^{\star})$. Choosing the initial and final state $\dd_0 = \dd_{final} = \dd_1$, $x_{\dd_0,0} = x_0 = z_{final}$ and switching times $t',t'',\ldots$ provides all needed to determine according to Algorithm 1 the trajectories of $z_{\dd}(t)$ and, thus, $u_{\dd}(t)$ from $\Psi_{\dd_i}$ (cf. Table \ref{tab:xplm1_di_muc}). Hence, the calculation of the system trajectories $\vekdd(k(t))$, $\dv(t)$, $z(t)$ and $u(t)$ is possible without integrating a differential equation and without solving a sequence search if $S$ and $P$ are pre--computed \footnote{Otherwise, fast online graph algorithms borrowed from computer science \cite{BaDeGoetal2016,DeSaScWa2009,WaWi2003} can be used.}. The simulation results are shown in Figure \ref{f:sim_tank}, for $l_0=5$, $c_{out}=0,\!5$, $c_{v,1}=0,\!8$, $c_{ovf}=0,\!2$, $l_1(t=0)=0,\!8$ and switching time intervals $t^{\star}=16$ ($[t]=\rm min$).

\subsection{Electrical network}
\label{ss:electr_network}

The second example is inspired by \cite{GeWoRuGue2006}, a work on flatness--based control of switched electrical circuits. Based on that application, an electrical DC network with two variable power sources $V_{in1}$ and $V_{in2}$ and two fluctuating loads $R_{L1}$ and $R_{L2}$ was modeled (Figure \ref{f:dc_elctr_ntwrk}). Two switches (controlled by the discrete inputs $\dv_1\in{0,1}$ and $\dv_2\in{0,1}$) allow to configure the network with increased or decreased damping and coupling properties. Aim is to control the voltage of load $1$ ($v_{L1}$) and the current of load $2$ ($i_{L2}$) by continuous inputs  $V_{in1}$ and $V_{in2}$. 

The switch positions of $\dv_1$ and $\dv_2$ yield four discrete states of a continuous system. It is assumed that for low load the switches are set to zero, i.e.

\ind{lll}{
if $v_{L1} < v_0$ & then & $\dv_1 = 0$, else $\dv_1 = 1$ \\[-0.5ex]
if $i_{L2} < i_0$ & then & $\dv_2 = 0$, else $\dv_2 = 1$ \,. \\[-1ex]
}

\begfig{9cm}
\psfrag{Vin1}{\small {$V_{in1}$}}
\psfrag{Vin2}{\small {$V_{in2}$}}
\psfrag{iL2}{{\small $i_{L2}$}}
\psfrag{i1}{{\small $i_{1}$}}
\psfrag{q1=0}{\small {$\dv_1=0$}}
\psfrag{q1=1}{\small {$\dv_1=1$}}
\psfrag{q2=0}{\small {$\dv_2=0$}}
\psfrag{q2=1}{\small {$\dv_2=1$}}
\psfrag{L}{\small {$L$}}
\psfrag{vC}{\small {$v_{C}$}}
\psfrag{vL1}{\small {$v_{L1}$}}
\psfrag{C}{\small {$C$}}
\psfrag{R}{\small {$R$}}
\psfrag{RL1}{\small {$R_{L1}$}}
\psfrag{RL2}{\small {$R_{L2}$}}
\efig{7cm}{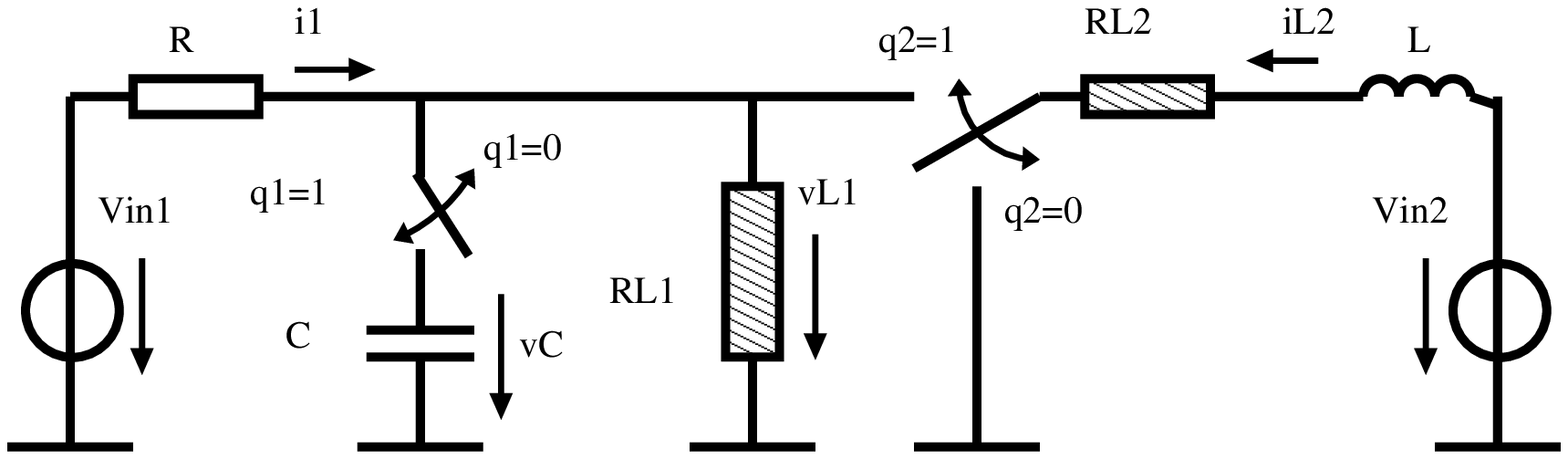}{DC electrical network}{f:dc_elctr_ntwrk}

Thereby, the capacitor is available to dampen step fluctuations of, e.g., $R_{L1}$ and $R_{L2}$, in case of higher network load. For the four discrete states, the continuous flat outputs are $z_1 = v_{L1}$ and $z_2 = i_{L2}$ and the continuous inputs are $u_1 = V_{in1}$ and $u_2 = V_{in2}$. Equation (\ref{eq:xmpl2_dv_di}) describes the dynamics of the system. \\[-7ex]

\be
\bega{rcl}
L \, \frac{d\,i_{L2}}{dt} & = & V_{in2} - ( R_{L2}\,i_{L2} + \dv_2\,v_{L1} ) \\
C \, \frac{d\,v_{C}}{dt} & = & \dv_1\,(i_1 - (\frac{1}{R_{L1}}\,v_{L1} - \dv_2\,i_{L2})) \\
\dv_1\,C \, \frac{d\,v_{C}}{dt} & = & i_1 - (\frac{1}{R_{L1}}\,v_{L1} - \dv_2\,i_{L2}) \\[-0.4ex]
V_{in1} & = & R\,i_1 + \dv_1\,v_C + (1-\dv_1)\,v_{L1} \\[-0.5ex]
\dv_1 \, v_C & = & \dv_1\,v_{L1} \,. \\[-2ex]
\ea
\label{eq:xmpl2_dv_di}
\ee

The initial condition for Equations (\ref{eq:xmpl2_dv_di}) is $v_C(t_0) = v_{C,0}$, $i_{L2}(t_0) = i_{L2,0}$. Permuting the discrete inputs $\dv_1$ and $\dv_2$ in Equations (\ref{eq:xmpl2_dv_di}) by their values $0,1$ yields the continuous system equations for the resepctive discrete states.\footnote{$[V]\!=\![v]\!=\!\rm V$, $[i]\!=\!\rm mA$, $[R]\!=\!\rm k\Omega$, $[C]\!=\!\rm F$, $[L]\!=\!\rm kH$} The continuous subsystems are flat, of which $\Psi_{\dd i}$ is given as follows:  \\[-4ex]

\ind{ll}{
$\dd_1$: & $\dv_1 = 0$, $\dv_2 = 0$ \\
{} & $u_1 = \left ( \frac{R}{R_{L1}} + 1 \right) \, z_1 $ \\
{} & $u_2 = L\,{\dot z}_2 + R_{L2} \, z_2 $ \\[-2.5ex]
}

\ind{ll}{
$\dd_2$: & $\dv_1 = 0$, $\dv_2 = 1$ \\
{} & $u_1 = \left ( \frac{R}{R_{L1}} + 1 \right) \, z_1 - R\, z_2$ \\
{} & $u_2 = L\,{\dot z}_2 + R_{L2} \, z_2 + z_1 $ \\[-2.5ex]
}

\ind{ll}{
$\dd_3$: & $\dv_1 = 1$, $\dv_2 = 0$ \\
{} & $u_1 = R\,C\,{\dot z}_1 + \left ( \frac{R}{R_{L1}} + 1 \right) \, z_1 $ \\
{} & $u_2 = L\,{\dot z}_2 + R_{L2} \, z_2 $ \\[-2.5ex]
}

\ind{ll}{
$\dd_4$: & $\dv_1 = 1$, $\dv_2 = 1$ \\
{} & $u_1 = R\,C\,{\dot z}_1 + \left ( \frac{R}{R_{L1}} + 1 \right) \, z_1 - R\,z_2 $ \\
{} & $u_2 = L\,{\dot z}_2 + R_{L2} \, z_2 + z_1 $ \\[-1ex]
}

The invariants and switching conditions are included in the adjacency list in Table \ref{tab:xplm2_AL}. The results of the trajectory planning according Algorithm 1 for path $P$ (Equation (\ref{eq:xmpl_P})) are shown in Figure \ref{f:sim_dc_ntwrk}, for $t^{\star}=16$ ($[t]=\rm s$) and \\[-4ex]

\ind{l}{
$R = 5$, $C=0,\!8$, $L=7$, $R_{L1}=2$, $R_{L2}=3$, \\
$v_0=6$, $i_0=0,\!5$, $v_{L1}(t_0) = 0,\!5$, $i_{L2}(t_0)=0,\!1$\,. \\[-2ex]
}

Like the one--tank example, the electrical network is a flat hybrid automaton. \\[-1ex]

\begfig{9cm}
\efig{9cm}{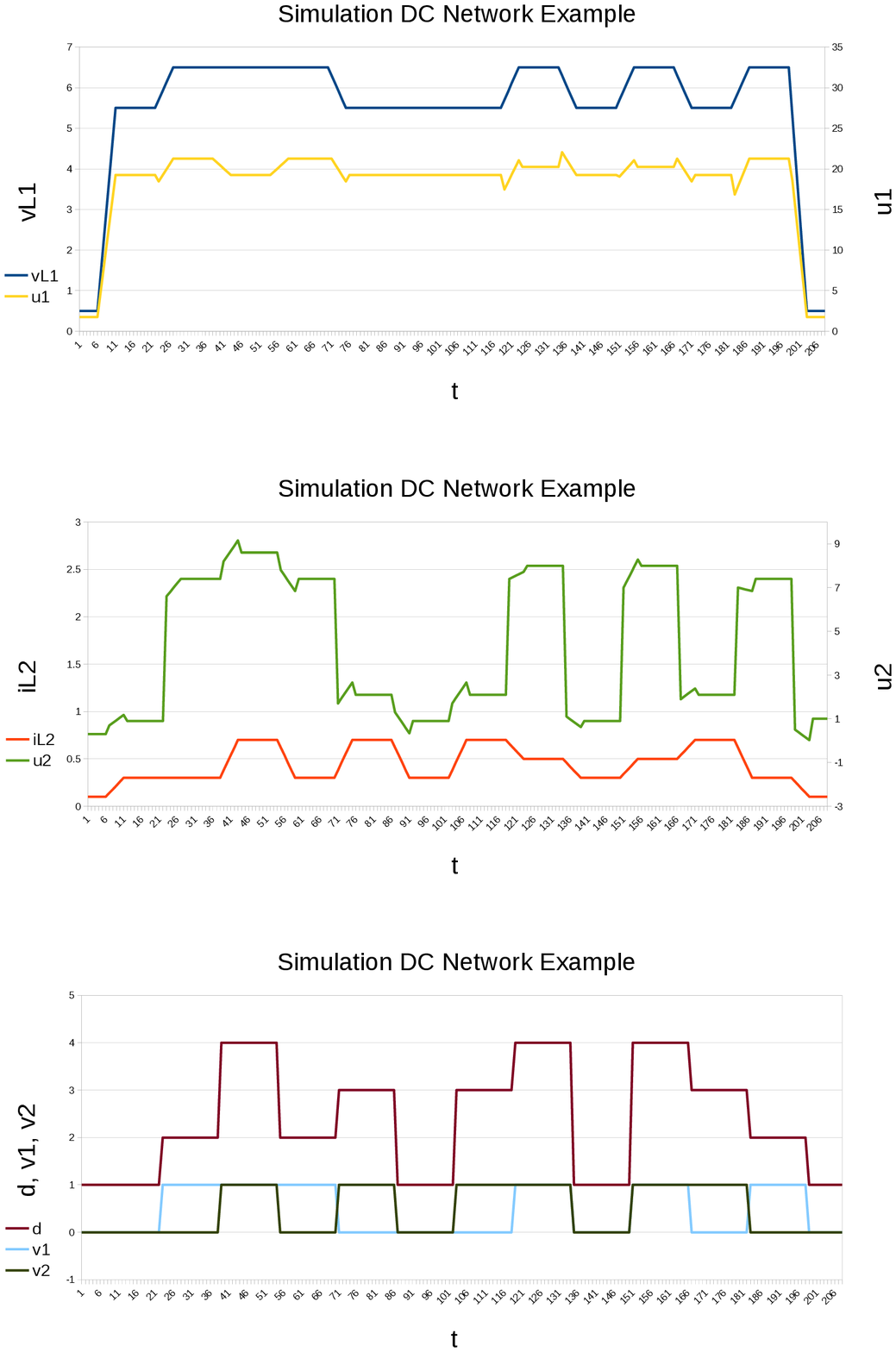}{Simulation of the DC Network example}{f:sim_dc_ntwrk}

{\bf Remark.} The electrical network example contains discrete--state transitions, for which several conditions have to be met simultaneously (transitions $\de_{3}$, $\de_{5}$, $\de_{8}$ and $\de_{10}$, cf. Table \ref{tab:xplm2_AL}). If, e.g., these transitions are removed, then $\rm A$ is still strongly connected and the system remains a $\rm FHA$. This shows that for a $\rm FHA$, a certain minimal realisation exists with the smallest number of discrete--state transitions.

\tabcap{htb}{cc|ccl}{
\hline
$\dd_i$ & $\mathcal{Z}^{\rm inv}_{\dd_i}\,,\; V^{\rm inv}_{\dd_i}$ & $\dd'_i$ & $\de_j$ & $\mathcal{G}^{\de_j}$ \\
\hline
$\dd_1$ & $v_{L1} < v_0$    & $\dd_2$ & $\de_1$ & $v_{L1} \geq v_0\,,\; \dv_1 = 1$ \\[-2ex]
{}      & $i_{L2} < i_0$    & $\dd_3$ & $\de_2$ & $i_{L2} \geq i_0\,,\; \dv_2 = 1$ \\[-2ex]
{}      & $\dv_1=0,\dv_2=0$ & $\dd_4$ & $\de_3$ & $v_{L1} \geq v_0\,,\; \dv_1 = 1$ \\[-2ex]
{}      & {}                & {}      & {}      & $i_{L2} \geq i_0\,,\; \dv_2 = 1$ \\[-0ex]
\hline
$\dd_2$ & $v_{L1} \geq v_0$ & $\dd_1$ & $\de_4$ & $v_{L1} < v_0\,,\; \dv_1 = 0$ \\[-2ex]
{}      & $i_{L2} < i_0$    & $\dd_3$ & $\de_5$ & $v_{L1} < v_0\,,\; \dv_1 = 0$ \\[-2ex]
{}      & $\dv_1=1,\dv_2=0$ & {}      & {}      & $i_{L2} \geq i_0\,,\; \dv_2 = 1$ \\[-2ex]
{}      & {}                & $\dd_4$ & $\de_6$ & $i_{L2} \geq i_0\,,\; \dv_2 = 1$ \\[-0ex]
\hline
$\dd_3$ & $v_{L1} < v_0$    & $\dd_1$ & $\de_7$ & $i_{L2} < i_0\,,\; \dv_1 = 0$ \\[-2ex]
{}      & $i_{L2} \geq i_0$ & $\dd_2$ & $\de_8$ & $v_{L1} \geq v_0\,,\; \dv_1 = 1$ \\[-2ex]
{}      & $\dv_1=0,\dv_2=1$ & {}      & {}      & $i_{L2} < i_0\,,\; \dv_2 = 0$ \\[-2ex]
{}      & {}                & $\dd_4$ & $\de_9$ & $v_{L1} \geq v_0\,,\; \dv_1 = 1$ \\[-0ex]
\hline
$\dd_4$ & $v_{L1} \geq v_0$ & $\dd_1$ & $\de_{10}$ & $v_{L1} < v_0\,,\; \dv_1 = 0$ \\[-2ex]
{}      & $i_{L2} \geq i_0$ & {}      & {}         & $i_{L2} < i_0\,,\; \dv_2 = 0$ \\[-2ex]
{}      & $\dv_1=1,\dv_2=1$ & $\dd_2$ & $\de_{11}$ & $i_{L2} < i_0\,,\; \dv_2 = 0$ \\[-2ex]
{}      & {}                & $\dd_3$ & $\de_{12}$ & $v_{L1} < v_0\,,\; \dv_1 = 0$ \\
}{8cm}{Adjacency list of the electrical network example}{tab:xplm2_AL}


\section{Conclusion and outlook}
\label{s:concl_outl}

The new class of Flat Hybrid Automata is introduced which allows to explicitly plan state and input trajectories from given output trajectories. Required system setup and properties are deduced, an approach for construction and for trajectory planning based on explicit system inversion is given and two demonstrative examples are discussed. Explicit input trajectory calculation can be especially of relevance if fast reaction for transition control is needed. Based on the FHA concept, design of explicitly schedulable networks with interconnected continuous systems that are switched on or off, respectively, can be approached. For these applications it may be necessary to design further inputs according to \cite{WaZe2008,WaZe2010} in order to obtain differentially flat continuous sub--systems. The solution is scalable in the sense that it is applicable to more complex systems, as long as the required properties are met or can be designed into the technical system, respectively. Not all possible discrete--state transitions must be considered to fulfil the requirements for a FHA. Hence, for future work it can be considered to realise a minimal Flat Hybrid Automaton with the least necessary number of state transitions for a given set of discrete states. Since the FHA can systematically be derived from a given state--space model of the considered dynamical system, it may be reasonable to develop an algorithmic approach for automatic deduction of the FHA. In case that the considered system is subject to uncertainties the question arises, how feedback control can be included into the feed--forward control of a FHA, based on e.g. \cite{HaDe2003siso,HaDe2003I,HaDe2003A,HaDe2010,HaZe2004}. The presented FHA concept and notation can be used in a theoretical context to further develop inversion and explicit input trajectory calculation of hybrid systems.

\bibliographystyle{plain}

\bibliography{paper_fha_bib}
\addcontentsline{toc}{section}{References}

\end{document}